\documentclass[12pt,preprint]{aastex}

\cpright{AAS}{2014}
\shorttitle{Active-Region Tilt Angles}
\shortauthors{Wang et al.}

\begin{document}

\title{Active-Region Tilt Angles: Magnetic Versus White-Light Determinations of Joy's Law}
\author{Y.-M. Wang and R. C. Colaninno}
\affil{Space Science Division, Naval Research Laboratory, Washington, DC 20375, USA}
\email{yi.wang@nrl.navy.mil, robin.colaninno@nrl.navy.mil}
\author{T. Baranyi}
\affil{Heliophysical Observatory, Research Centre for Astronomy and Earth Sciences, Hungarian Academy of Sciences, 4010 Debrecen, Hungary}
\email{baranyi@tigris.unideb.hu}
\and
\author{J. Li}
\affil{Department of Earth, Planetary, and Space Sciences, UCLA, Los Angeles, CA 90095, USA}
\email{jli@igpp.ucla.edu}

\begin{abstract}
The axes of solar active regions are inclined relative to the east--west 
direction, with the tilt angle tending to increase with latitude 
(``Joy's law'').  Observational determinations of Joy's law have been 
based either on white-light images of sunspot groups or on magnetograms, 
where the latter have the advantage of measuring directly the 
physically relevant quantity (the photospheric field), but the 
disadvantage of having been recorded routinely only since the mid-1960s.  
White-light studies employing the historical Mount Wilson (MW) database 
have yielded tilt angles that are smaller and that increase less steeply 
with latitude than those obtained from magnetic data.  We confirm 
this effect by comparing sunspot-group tilt angles from the 
Debrecen Photoheliographic Database with measurements made by 
Li and Ulrich using MW magnetograms taken during cycles 21--23.  
Whether white-light or magnetic data are employed, the median tilt angles 
significantly exceed the mean values, and provide a better characterization 
of the observed distributions.  The discrepancy between the 
white-light and magnetic results is found to have two main sources.  
First, a substantial fraction of the white-light ``tilt angles'' 
refer to sunspots of the same polarity.  Of greater physical significance 
is that the magnetograph measurements include the contribution 
of plage areas, which are invisible in white-light images 
but tend to have greater axial inclinations than the adjacent sunspots.  
Given the large uncertainties inherent in both the white-light 
and the magnetic measurements, it remains unclear whether any 
systematic relationship exists between tilt angle and cycle amplitude 
during cycles 16--23.
\end{abstract}


\section{INTRODUCTION}
As toroidal magnetic flux rises buoyantly through the convection zone 
to form sunspots and active regions, the Coriolis force acts to twist 
the flux tubes so that they emerge slightly inclined relative to 
the east--west direction (see, e.g., Caligari et al. 1995; 
D'Silva \& Choudhuri 1993; Fan et al. 1994; Fisher et al. 1995; 
Wang \& Sheeley 1991).  This axial tilt plays a central part in 
flux-transport dynamo models, where it is an essential ingredient 
in the formation and evolution of the polar fields 
(see, e.g., Wang \& Sheeley 1991; Dikpati \& Charbonneau 1999; 
Mackay et al. 2002; Baumann et al. 2004; Cameron et al. 2010; 
Mu{\~n}oz-Jaramillo et al. 2010, 2013; Jiang et al. 2011a, 2011b, 2013, 2014; 
Kitchatinov \& Olemskoy 2011; Cameron \& Sch{\"u}ssler 2012; 
Hathaway \& Upton 2014; Upton \& Hathaway 2014a, 2014b).

The tendency for the leading/westward parts of sunspot groups to be 
located equatorward of their trailing/eastward parts was first studied 
by Joy, using the sunspot drawings of Carrington and Sp{\"o}rer 
made between 1856 and 1893 (cycles~10--13).  From the data listed 
in Table~II of Hale et al. (1919), we infer that the 2633 sunspot groups 
measured by Joy had an area-weighted average inclination of 6{\fdg}1; 
a linear fit to the tilt angles as a function of latitude $L$ gives
\begin{equation}
\gamma(L) = 0.27\vert L\vert + 1\fdg9
\end{equation}
over the range $0 < \vert L\vert < 30^\circ$.  By definition,
\begin{equation}
\tan\gamma = \frac{\Delta L}{\Delta\phi\cos L},
\end{equation}
where $\Delta L$ is the latitudinal separation between the leading and 
trailing poles of the sunspot group or active region, $\Delta\phi$ 
is their longitudinal separation, and $\gamma$ is taken to be positive 
if the leading pole lies equatorward of the trailing pole (in either 
hemisphere), negative otherwise.  This definition restricts the 
tilt angles to the range $-90^\circ\leq\gamma\leq +90^\circ$, 
and does not distinguish between different polarity orientations, 
so that it is suited for white-light measurements in which the polarities 
are unknown.  Hale et al. (1919) noted that ``the angle of inclination 
was found to depend entirely on the latitude of the group, without 
reference to the number of the cycle or the time within the cycle.  
A knowledge of the polarities of the spots would have aided greatly 
in determining the position of the axes of the groups.''

Joy's results, as reported in Hale et al. (1919), were subsequently 
confirmed by Brunner (1930), using drawings of 1981 sunspot groups 
made at the Z{\"u}rich Observatory between 1894 and 1928 (cycles~13--16).  
Brunner obtained a value of $\gamma_{\rm mean} = 6{\fdg}5$ for the 
mean tilt angle of his entire sample; a linear fit to the latitudinally 
binned data in his Table~I yields
\begin{equation}
\gamma(L) = 0.51\vert L\vert - 0{\fdg}8,
\end{equation}
the slope thus being considerably steeper than that found by Joy.  
It should be noted that the derived slopes depend sensitively on the 
relatively few measurements made at very low and very high latitudes: 
thus, for the 0$^\circ$--5$^\circ$ (25$^\circ$--30$^\circ$) latitude bin, 
Brunner obtained a mean tilt angle of 0{\fdg}6 (14{\fdg}4), 
whereas Joy's measurements gave 3{\fdg}7 (9{\fdg}3), leading to 
a much flatter slope.  According to Brunner, the average axial tilts 
of sunspot groups decrease systematically from $\sim$7{\fdg}8 
to $\sim$4{\fdg}6, as they evolve from small spots/pores without penumbrae 
to large, widely separated spots.  In contrast, Hale et al. (1919) 
stated that ``there is little change in this angle during the life 
of the group.''

Joy's law was confirmed using magnetograph measurements by 
Wang \& Sheeley (1989, 1991), who analyzed the statistical properties 
of 2710 bipolar magnetic regions (BMRs) that appeared in daily full-disk 
magnetograms recorded by the National Solar Observatory at Kitt~Peak 
(NSO/KP) during cycle~21 (1976--1986).  Here, the centroidal positions 
of the positive- and negative-polarity flux were estimated visually by 
Sheeley from photographic prints, with the help of Stonyhurst overlay grids.  
The BMR tilt angles were found to increase with latitude with a slope 
of approximately 0.5, close to the value obtained by Brunner (1930); 
averaged over the entire data set, $\gamma_{\rm mean}$ was as large as 
9$^\circ$.  An important difference from most other tilt-angle studies 
is that each BMR was measured only once, at or as close as possible 
to the time when its flux reached its peak (the original purpose 
having been to provide a source term for surface flux-transport modeling).  
By tracking the day-to-day evolution of the flux by eye, it was 
usually possible to distinguish between closely spaced active regions, 
which is often not the case when automated methods are employed.

Howard (1989, 1991a) applied a field-strength thresholding algorithm 
to identify active regions in coarse-format versions of Mount Wilson (MW) 
Observatory daily magnetograms recorded during 1967--1990, 
and then derived their axial tilts from the flux-weighted positions 
of their leading and trailing sectors.  He obtained a less regular increase 
in the tilt angle with latitude than Wang \& Sheeley (1989), 
and suggested that the difference might be due to the inclusion of 
older, decayed regions in his sample.  

From an analysis of 517 BMRs appearing on magnetograms taken 
at the Huairou Solar Observing Station during 1988--2001, 
Tian et al. (2003) derived a Joy's law slope of 0.5, 
in good agreement with the results of Wang \& Sheeley (1989, 1991).  
The active regions in this study were selected for measurement 
when they were ``almost mature'' and located within 15$^\circ$ 
of central meridian.

More recently, Stenflo \& Kosovichev (2012) and Li \& Ulrich (2012) 
developed automated procedures to derive active-region tilt angles 
from magnetograms recorded with the Michelson Doppler Imager (MDI) 
on the {\it Solar and Heliospheric Observatory} ({\it SOHO}), 
finding slopes of $\sim$0.56 and 0.36, respectively, for Joy's law.  
Li \& Ulrich (2012) also applied their algorithm to MW magnetograms 
taken during 1974--2012, obtaining a slope of 0.45 and a mean tilt angle 
of 6{\fdg}2.

Howard (1991b) derived tilt angles for sunspot groups appearing 
in MW daily white-light photographs taken between 1917 and 
1985.\footnote{See \texttt{ftp://ftp.ngdc.noaa.gov/STP/SOLAR\_DATA/
SUNSPOT\_REGIONS/Mt\_Wilson\_Tilt}.}  Here, the umbral positions and 
areas were measured and grouped by proximity as described in 
Howard et al. (1984); the umbrae located westward (eastward) of the 
area-weighted centroid of each group were then defined as belonging 
to the ``leading'' (``following'') sector (despite the absence of 
polarity information).  As may be seen from Figure~3 in Howard (1991b), 
the white-light tilt angles show a fairly regular increase 
up to latitude $\vert L\vert\sim 30^\circ$, but the average slope 
is only about one-half that shown in Figure~5 of Howard (1991a) 
for the tilts derived from MW magnetograms.  Reviewing his white-light 
and magnetic results, Howard (1996) noted that the average tilt angle 
based on the 24,308 MW sunspot-group measurements over the 
period 1917--1985 was $\gamma_{\rm mean} = 4{\fdg}3$, whereas 
the corresponding value for the 15,692 ``plages'' in MW magnetograms 
taken during 1967--1995 was $\gamma_{\rm mean} = 6{\fdg}3$.

Employing the MW sunspot-group tilt-angle measurements and 
the analogous Kodaikanal (KK) database covering the period 1906--1987, 
Dasi-Espuig et al. (2010) derived a slope of 0.26/0.28 (MW/KK) for Joy's law 
and a mean tilt angle of 4{\fdg}25/4{\fdg}5 (MW/KK).  In addition, 
they found a tendency for $\gamma_{\rm mean}/\vert L\vert_{\rm mean}$, 
the mean tilt angle normalized by the mean latitude of the sunspot groups, 
calculated for each of cycles~15--21, to decrease with increasing 
cycle strength.  From the Pulkovo database of sunspot group measurements 
during 1948--1991, Ivanov (2012) obtained a Joy's law slope of 0.38 
and a mean tilt angle of 6{\fdg}1.  He also showed that the anticorrelation 
between $\gamma_{\rm mean}/\vert L\vert_{\rm mean}$ and cycle strength 
in the MW data is almost entirely due to the unusually low value of 
$\gamma_{\rm mean}$ during cycle~19 (see also the corrigendum of 
Dasi-Espuig et al. 2013; a strong anticorrelation remains present 
in the normalized KK tilt angles for cycles~15--21).  In their study 
of hemispheric asymmetries in the MW sunspot-group tilt angles, 
McClintock \& Norton (2013) found that the mean value of 
$\gamma$ during cycle~19 was 4{\fdg}6 in the northern hemisphere, 
but only 1{\fdg}8 in the southern hemisphere.  In line with the results 
of Dasi-Espuig et al. (2010) and the flux transport simulations 
of Sch{\"u}ssler \& Baumann (2006), they also advocated a 
downward revision in the slope of Joy's law.

From this brief review, we infer that the active-region tilt angles 
derived from magnetograms have tended to be systematically larger than 
those deduced from white-light observations.  As an apparent exception, 
the sunspot group measurements of Brunner (1930) are in accord with the 
magnetograph results.  However, an inspection of Brunner's Table~III 
suggests that roughly one-third of his measurements were made at a 
very early, pre-penumbral stage in the evolution of the sunspot groups, 
when their mean tilt was as large as 7{\fdg}8.  Such emerging regions, 
which consist of small, faint sunspot pores, are likely to constitute 
a much smaller fraction of the sunspot groups in the MW and KK 
databases, which are presumably dominated by fully developed regions 
whose tilt angles have an average value of 4{\fdg}6, according to 
Table~III of Brunner (1930).  Since the statistical studies of 
magnetic tilt angles are also heavily weighted toward fully emerged 
active regions, the good agreement between Brunner's results and those 
of Wang \& Sheeley (1989, 1991), Tian et al. (2003), Li \& Ulrich (2012), 
and Stenflo \& Kosovichev (2012) could be partly fortuitous.

The main objective of this study is to understand why white-light 
observations of sunspot groups tend to yield smaller tilt angles 
than magnetograph measurements of active regions.  For this purpose, 
we compare the tilt angles given in the Debrecen Photoheliographic Data (DPD) 
sunspot catalog\footnote{See \texttt{http://fenyi.solarobs.unideb.hu/DPD}.  
The DPD database represents a continuation of the Greenwich 
Photoheliographic Results (1874--1976).} with the corresponding values 
derived by Li \& Ulrich (2012) from MW magnetograms taken during 
cycles~21--23 (Section~2).  Section~3 places these recent magnetic 
and white-light tilt-angle measurements in the context of the 
MW white-light results for cycles~16--21.  In Section~4, we examine 
specific cases in which the white-light and magnetic measurements 
give very different tilt angles, and discuss how such systematic 
differences arise.  Our results are summarized in Section~5.

\section{COMPARISON OF WHITE-LIGHT AND MAGNETIC TILT ANGLES DURING CYCLES 21--23}
The online DPD catalog contains daily measurements of sunspot areas and 
positions from 1974 to the present, based on full-disk white-light images 
taken mainly at the Debrecen Observatory and its Gyula Observing Station 
(see Gy\H{o}ri et al. 2011).  The database also provides sunspot-group 
tilt angles derived in two different ways: (1) from the area-weighted 
umbral positions alone (``U''), as in the case of the MW and KK 
sunspot-group tilt angles; and (2) from the area-weighted positions 
of all visible components of the sunspot group, including umbrae, penumbrae, 
and pores (``UPP'').\footnote{The DPD catalog designates the combination 
of umbral and non-umbral areas by ``whole spot'' or ``WS'' (formerly 
``U+P'').  However, to emphasize that the non-umbral areas include 
small pores as well as penumbrae, we here employ the acronym ``UPP'' 
(umbrae, penumbrae, and pores) instead of ``WS.''}  Again, 
the ``leading'' and ``following/trailing'' portions of a sunspot group 
are defined relative to the area-weighted centroid of the group, 
and do not necessarily coincide with the actual leading- and 
trailing-polarity sectors of the active region.  We also remark that 
the number of DPD/UPP tilt-angle measurements considerably exceeds 
(by $\sim$70\%) the number of DPD/U measurements, because many 
sunspot groups contain at least a few faint pores, but only one umbra 
(usually in the leading sector) or, in a roughly equal number of cases, 
no umbrae at all.

In addition to DPD, the Debrecen Observatory team has constructed 
a database that employs MDI continuum images and magnetograms recorded 
during 1996--2011.  The {\it SOHO}/MDI Debrecen Data (SDD) sunspot catalog 
provides four kinds of tilt angle measurements, two of which use only 
the continuum intensity images and are exactly analogous to DPD/U 
and DPD/UPP.  The remaining two types of tilt angle estimates 
exploit the MDI magnetograms to assign polarities to the U and UPP 
continuum measurements, so that the leading and following sectors 
are distinguished by polarity instead of being defined by the 
area-weighted centroid of the sunspot group.  Because the 
statistical properties of the SDD and DPD tilt angles turn out 
to be rather similar, however, this study will focus on the DPD database, 
which covers three complete sunspot cycles.  A detailed analysis 
of the SDD and DPD measurements and comparison with MW/KK may be 
found in Baranyi (2014).

Employing daily full-disk MW magnetograms recorded between 1974 
and 2012, Li \& Ulrich (2012) determined the tilt angles of active regions
based on the flux-weighted centroids of the leading- and 
trailing-polarity sectors.  Since the period that they studied 
nearly coincides with that covered by the DPD database, it is interesting 
to compare their magnetic tilt angles (``MW/MAG'') with the 
DPD white-light results.  

Let the tilt angles derived from the DPD/U, DPD/UPP, and MW/MAG 
measurements be designated by $\gamma_{\rm U}$, $\gamma_{\rm UPP}$, 
and $\gamma_{\rm mag}$, respectively.  Throughout this study, 
we include only measurements made within 45$^\circ$ of longitude 
from central meridian, so as to avoid distortions due to foreshortening 
near the solar limb.  To determine the latitudinal dependence of the 
tilt angles, we bin the measurements into 5$^\circ$-wide intervals 
of unsigned latitude $\vert L\vert$, and calculate either the 
arithmetic mean or the median of the data within each bin.  
When performing averages, we follow the convention of treating 
multiple measurements of a single active region as if they represented 
measurements of different active regions (instead of combining them 
into one value), even though this procedure gives greater weight 
to longer-lived regions.

We begin by examining the average behavior of the tilt angles 
over the entire interval from 1976 June through 2008 December, comprising 
cycles~21--23.  For this 32~yr period, Li \& Ulrich (2012) made a total 
of 25,322 measurements of $\gamma_{\rm mag}$ based on 7958 active regions; 
the DPD database provides 23,204 measurements of $\gamma_{\rm U}$ 
for 5973 NOAA sunspot groups, and 38,975 measurements of $\gamma_{\rm UPP}$ 
for 8575 NOAA sunspot groups.  (Although the DPD catalog also contains 
sunspot groups without NOAA designations, we omit them in order to 
maximize the overlap with the active region sample of Li and Ulrich.)  
Figure~1 shows $\gamma_{\rm mag}$, $\gamma_{\rm U}$, and $\gamma_{\rm UPP}$ 
as a function of $\vert L\vert$.  In the top panel, arithmetic means 
of the tilt angles are plotted for each latitude bin between 
0$^\circ$--5$^\circ$ and 30$^\circ$--35$^\circ$, while median values 
are plotted in the bottom panel.  The standard error of the mean/median 
is indicated by the vertical bar through each point; the errors 
are largest for the highest- and lowest-latitude bins because they contain 
the smallest number of measurements.  The straight lines 
represent least-squares fits to the plotted points up to latitude 
$\vert L\vert = 30^\circ$; the point at 30$^\circ$--35$^\circ$ 
is omitted both because it is based on relatively few measurements, 
and because it may be significantly affected by differential rotation, 
which acts to decrease the tilt angles with time and has its 
steepest gradients at midlatitudes.  The slope $m_{\rm Joy}$ and 
$y$-intercept $b_0$ of the regression line then define Joy's law 
for the given set of tilt angle measurements:
\begin{equation}
\gamma(L) = m_{\rm Joy}\vert L\vert + b_0.
\end{equation}

From Figure 1, the magnetic tilt angles ($\gamma_{\rm mag}$) are seen 
to increase more steeply than those derived from white-light observations 
($\gamma_{\rm U}$, $\gamma_{\rm UPP}$).  In addition, regardless of 
whether magnetic or white-light data are used, the median values 
yield systematically larger slopes $m_{\rm Joy}$ than the mean values.  

If the tilt angle measurements for all latitudes are now combined, 
we find that the mean values of $\gamma_{\rm mag}$, $\gamma_{\rm U}$, 
and $\gamma_{\rm UPP}$ during cycles~21--23 are $6{\fdg}2\pm 0{\fdg}2$, 
$5{\fdg}3\pm 0{\fdg}2$, and $5{\fdg}9\pm 0{\fdg}1$, respectively, while 
their median values are $7{\fdg}5\pm 0{\fdg}2$, $6{\fdg}6\pm 0{\fdg}2$, 
and $6{\fdg}9\pm 0{\fdg}2$ (all quoted errors are 1$\sigma$).  
Thus the magnetic tilt angles are systematically larger than their 
white-light counterparts, and the median values are systematically larger 
than the mean values for both magnetic and white-light tilts.  
The white-light tilt angles derived from umbral, penumbral, and 
pore positions are slightly larger than those derived from umbral 
positions only.

Both magnetic and white-light tilt angles exhibit an extremely wide scatter, 
due both to measurement errors and to the randomizing effect of the 
ambient supergranular convection; in general, the root-mean-square (rms) 
tilt angles are much larger than their arithmetic means, and increase 
with decreasing BMR size or total flux (see Howard 1989, 1991a, 1991b, 1996; 
Wang \& Sheeley 1989, 1991).  Figure~2 shows histograms of $\gamma_{\rm mag}$, 
$\gamma_{\rm U}$, and $\gamma_{\rm UPP}$, again including all measurements 
taken during cycles~21--23.  The bin width is 5$^\circ$, and each histogram 
(plotted as a smooth curve connecting the discrete bin values) has been 
normalized by dividing by the total number of measurements and multiplying 
by a constant scaling factor.  All three tilt-angle distributions exhibit 
a Gaussian-like profile, with peak located in the 5$^\circ$--10$^\circ$ 
bin and a half width at half-maximum (HWHM) ranging from 15{\fdg}4 (DPD/U) 
to 17{\fdg}5 (MW/MAG).  (Despite their Gaussian appearance, however, 
the distributions are not well fitted by single Gaussian functions.)  
Because the peak is shifted to the right of $\gamma = 0$, 
the wings of the profile are not symmetric over the range 
$-90^\circ < \gamma < +90^\circ$, and the ``outliers'' 
near the left/negative end of the distribution act to decrease 
the mean value of $\gamma$ relative to the median value.  To represent 
the centroid of such skewed distributions, it is preferable to employ 
the median (as was done by Li \& Ulrich 2012) rather than the mean.  
From a physical viewpoint, we note that a nonnegligible fraction of 
the negative outliers are likely to represent BMRs with 
reversed (``anti-Hale'') east--west polarity orientations, which would 
be assigned tilt angles $\gamma > +90^\circ$ if a polarity-dependent 
definition were adopted, in which $\gamma$ ranges between $-$180$^\circ$ 
and $+$180$^\circ$.  In the absence of the artificial cutoff 
at $\pm$90$^\circ$, the mean of the distribution would then increase 
and converge toward the median.  It should be remarked that 
the great majority of extreme outliers beyond 
$\vert\gamma\vert\sim 60^\circ$ are associated with small active regions 
having leading--following separations of less than 2{\fdg}5.

Although the distributions of $\gamma_{\rm mag}$, $\gamma_{\rm U}$, and 
$\gamma_{\rm UPP}$ have remarkably similar, quasi-Gaussian profiles, 
the curves of Figure~2 deviate significantly from each other 
around their peaks and for relatively large, positive tilt angles.  
In particular, MW/MAG shows a ``deficit'' of tilt angles in the range 
$\sim$0$^\circ$--15$^\circ$ but a ``surplus'' in the range 
$\sim$25$^\circ$--40$^\circ$, as compared to DPD/UPP and especially DPD/U.  
From the fact that tilt angles of $\sim$25$^\circ$--40$^\circ$ 
are characteristic of active regions located at mid to high latitudes, 
we infer that such active regions (typically associated with the 
rising phase of the cycle) may be an important source of the differences 
among the three distributions, as is also suggested by Figure~1.

We now consider the behavior of the tilt angles for the individual cycles.  
In the top panel of Figure~3, we plot the median values of $\gamma_{\rm mag}$ 
as a function of $\vert L\vert$ separately for cycles~21, 22, and 23; 
the middle and bottom panels display the corresponding results for 
$\gamma_{\rm U}$ and $\gamma_{\rm UPP}$, respectively.  Neither the 
magnetic nor the white-light tilt angles show any clear evidence 
for systematic variations among these three cycles (which, 
it should be remarked, have rather similar sunspot-number amplitudes).  
In the case of the tilt angles derived from MW magnetograms, 
the slope $m_{\rm Joy}^{\rm (med)}$ ranges from $0.47\pm 0.05$ (cycle~22) 
to $0.57\pm 0.03$ (cycle~21); the median tilt angle based on all latitudes, 
$\gamma_{\rm med}$, ranges from $7{\fdg}1\pm 0{\fdg}3$ (cycle~23) 
to $7{\fdg}8\pm 0{\fdg}4$ (cycle~22).  In the case of the tilt angles 
derived from umbral data alone, $m_{\rm Joy}^{\rm (med)}$ varies from 
$0.34\pm 0.02$ (cycle~21) to $0.46\pm 0.06$ (cycle~22), while 
$\gamma_{\rm med}$ varies from $6{\fdg}0\pm 0{\fdg}4$ (cycle~22) to 
$7{\fdg}0\pm 0{\fdg}4$ (cycle~23).  When penumbral areas and pores are 
included, $m_{\rm Joy}^{\rm (med)}$ varies from $0.36\pm 0.03$ (cycle~21) 
to $0.44\pm 0.04$ (cycle~23), and $\gamma_{\rm med}$ varies 
from $6{\fdg}5\pm 0{\fdg}3$ (cycle~22) to $7{\fdg}4\pm 0{\fdg}3$ (cycle~23).  
Thus $m_{\rm Joy}^{\rm (med)}$ has its largest value in cycle~21 
according to the magnetic measurements, in cycle~22 when umbral positions 
alone are used, but in cycle~23 when umbral, penumbral, and pore data 
are employed; on the other hand, $\gamma_{\rm med}$ has its 
largest value in cycle~22 according to the magnetic observations, 
but in cycle~23 according to both sets of white-light observations.  
These seemingly random differences give some indication of the uncertainties 
involved in attempting to deduce systematic long-term trends from 
tilt angle measurements.

The tilt angle statistics described here and in the next section 
are summarized in Tables~1 and 2.

\section{COMPARISON WITH THE MW SUNSPOT-GROUP DATABASE}
The tilt angles derived from the sunspot-group measurements of 
Howard et al. (1984), using daily white-light photographs taken at MW 
during 1917--1985, have been analyzed previously by Howard (1991b, 1996), 
Dasi-Espuig et al. (2010), Ivanov (2012), and McClintock \& Norton (2013).  
Because Howard et al. (1984) measured the positions and areas of umbrae only, 
this data set, referred to here as ``MW/WL,'' may be considered analogous 
to DPD/U.  It is clearly of interest to compare the MW/WL tilt angles 
with those derived from the more recent databases.  For this purpose, 
we divide the MW/WL time series into six cycles, starting with cycle~16 
(1923~August--1933~August) and ending with cycle~21 (1976~June--1986~August); 
we discard the observations for cycle~15, since they do not include its 
rising phase, and we assume that the data gap during the final months 
of cycle~21 does not significantly affect the overall statistics for that 
cycle.  We omit all measurements taken more than 45$^\circ$ from 
central meridian longitude, as well as those for which the centroids 
of the leading and following parts are separated by an angle 
$\Delta s = [(\Delta\phi\cos L)^2 + (\Delta L)^2]^{1/2}$ exceeding 
20$^\circ$, which is the maximum pole separation among the BMRs 
analyzed by Wang \& Sheeley (1989; see their Figure~2, and also the 
discussion of Dasi-Espuig et al. 2010).  To derive Joy's law 
for each cycle, we again compute the mean or median value of 
$\gamma$ within 5$^\circ$-wide bins in $\vert L\vert$ and apply a 
linear least-squares fit over the latitude range 0$^\circ$--30$^\circ$.

Figure 4 (top panel) shows the Joy's law slope, $m_{\rm Joy}$, derived 
for each of cycles~16--21 by fitting the mean values of $\gamma$ as a 
function of $\vert L\vert$.  Also plotted for comparison are the 
corresponding slopes obtained for cycles~21--23 from the 
DPD/U, DPD/UPP, and MW/MAG measurements.  As a reminder of 
the relative strengths of the different cycles, the asterisks indicate 
the (arbitrarily scaled) maximum yearly sunspot number $R_{\rm max}$ 
for each cycle.  The bottom panel of Figure~4 displays the same quantities, 
but with $m_{\rm Joy}$ determined from the median rather than 
the mean value of the tilt angle as a function of latitude.

From Figure 4, we see that the Joy's law slopes obtained from 
the MW/WL sunspot-group data tend to be substantially smaller than 
those determined from both the DPD and the magnetic measurements.  
Averaged over cycles~16--21, the mean-based (median-based) MW/WL slope 
is $m_{\rm Joy} = 0.20\pm 0.01$ (0.27$\pm$0.01),\footnote{The value 
of $m_{\rm Joy}^{(\rm mean)}$ derived by Dasi-Espuig et al. (2010) 
from the MW/WL data (0.26) is higher than ours (0.20) because they forced 
their least-squares fit to pass through zero, setting $b_0 = 0$ 
in Equation~(4) (see also the discussion of McClintock \& Norton 2013).} 
whereas the corresponding value for DPD/U, averaged over cycles~21--23, 
is $m_{\rm Joy} = 0.28\pm 0.03$ (0.40$\pm$0.04).  However, for cycle~21, 
the only period when the databases overlap, MW/WL actually gives a 
larger slope than DPD/U; the significance of this ``convergence'' 
is unclear to us.

Figure 5 (top panel) displays the mean values of $\gamma$, averaged 
over each cycle and over all latitudes, for the different data sets; 
the median values are plotted in the bottom panel.  When averaged over 
cycles~16--21, the MW/WL tilt angles have a mean (median) value of 
4{\fdg}4$\pm$0{\fdg}2 (5{\fdg}8$\pm$0{\fdg}2), roughly $1^\circ$ smaller 
than the corresponding DPD values averaged over cycles~21--23, 
and $\sim$2$^\circ$ smaller than the MW/MAG values.  Again, 
the different data sets are in relatively good agreement for cycle~21, 
when the MW/WL tilt angles attain their highest cycle-averaged values.  
We note that the DPD tilt angles for cycle~21 remain almost unchanged 
even if the 1986 measurements are omitted when averaging over this cycle; 
thus it is unlikely that the agreement can be attributed to the 
absence of MW/WL data in 1986.

As pointed out by Dasi-Espuig et al. (2010), the cycle-averaged values 
of $\gamma$ depend on the latitudinal distribution of the 
measured active regions, which may vary from cycle to cycle 
and from one database to another.  To correct for this effect 
(following Dasi-Espuig et al.), we now divide the cycle-averaged tilt angles 
by their corresponding cycle-averaged, unsigned latitudes.  
Figure~6 shows the resulting normalized quantities, 
$\gamma_{\rm mean}/\vert L\vert_{\rm mean}$ (top panel) and 
$\gamma_{\rm med}/\vert L\vert_{\rm med}$ (bottom panel).  
The latitude normalization considerably reduces the cycle-to-cycle 
variations and the differences between the various data sets.  
This partial convergence reflects the fact that, averaged over 
cycles~16--21, the MW/WL measurements were made at a mean (median) latitude 
of 15{\fdg}1 (14{\fdg}4), as compared to 15{\fdg}9 (15{\fdg}3) 
for the DPD measurements, averaged over cycles~21--23.

Even after the latitude normalization is applied, it is clear 
from Figure~6 that cycle~19 remains an outlier characterized by 
unusually small values of $\gamma_{\rm mean}/\vert L\vert_{\rm mean}$ 
and $\gamma_{\rm med}/\vert L\vert_{\rm med}$, according to the 
MW/WL measurements.  For comparison, as indicated by Figure~4(c) 
in Ivanov (2012), the Pulkovo measurements for this cycle give 
$\gamma_{\rm mean}/\vert L\vert_{\rm mean}\simeq 0.40$, 
twice the MW/WL value.  However, it should also be noted that 
the Pulkovo data coverage of cycle~19 is incomplete, with Figure~3 
of Ivanov (2012) showing a gap that extends through the maximum phase 
1957--1960.  As pointed out by McClintock \& Norton (2013), 
the unusually low MW/WL value for cycle~19 is mainly attributable 
to the southern-hemisphere sunspot groups, which have a 
latitude-normalized mean tilt angle of 0.15, as compared to 0.26 
for the northern hemisphere.  In flux transport models, 
the evolution of the polar fields depends sensitively on the tilt angles 
of active regions and on the meridional flow speed.  The magnetograph 
observations of Babcock (1959) indicate that the south polar field 
reversed as early as mid-1957, whereas the north polar field reversed 
in late 1958 (see also the polar faculae counts of Sheeley 2008).  
This timing seems difficult to reconcile with the very small tilt angles 
that the MW/WL database assigns to the southern hemisphere, 
unless large variations in the meridional flow speed are invoked.

In Figure 7, we have plotted a histogram of the 22,533 MW/WL tilt-angle 
measurements made during cycles~16--21; for comparison, we again display 
the histogram of the 23,204 DPD/U tilt angles measured during cycles~21--23.  
The MW/WL tilt angles show a significantly larger scatter, with the 
Gaussian-like distribution having a HWHM of 17{\fdg}9, as compared to 
15{\fdg}4 for the DPD/U measurements.

From a sample of MW white-light images and sunspot-group drawings 
with polarity labeling\footnote{See 
\texttt{ftp://howard.astro.ucla.edu/pub/obs/drawings}.  The extensive 
information contained in these drawings, which identify each sunspot group 
and indicate the positions, areas, polarities, and strengths of 
most of the individual sunspots/pores recorded since 1917, has recently 
been tabulated by Tlatov et al. (2014).} for the period 1961--1967, 
Baranyi (2014) found that $\sim$84\% of the sunspot groups 
whose leading and following parts had separations 
$\Delta s < 2{\fdg}5$ consisted entirely of spots of the same polarity.  
In the great majority of these ``unipolar'' groups, the actual 
opposite-polarity (usually trailing) sector contained no visible umbrae; 
in the remaining cases, the two sectors of a larger sunspot group 
were incorrectly taken to be separate groups.  Based on all of the 
MW/WL measurements made within 45$^\circ$ of central meridian 
during cycles~16--21, the median angular separation of the 
leading and following sectors is only 2{\fdg}9, and 43\% of the 
sunspot groups have $\Delta s < 2{\fdg}5$.  As may be seen from 
the red dotted curve in Figure~8, the tilt angle distribution of these 
``compact'' groups is extremely broad (HWHM = 25{\fdg}7); furthermore, 
the mean (median) value of $\gamma$ is only 2{\fdg}7 (4{\fdg}0), 
as compared to 4{\fdg}4 (5{\fdg}8) for all sunspot groups.  
When this component is separated out and only sunspot groups having 
$\Delta s > 2{\fdg}5$ are retained, the HWHM of the original tilt-angle 
distribution decreases from 17{\fdg}9 to 16{\fdg}6, and the outliers 
in the far wings $\vert\gamma\vert\gtrsim 60^\circ$ almost disappear 
(compare the black solid and blue dashed curves in Figure~8).

After filtering out sunspot groups with $\Delta s < 2{\fdg}5$ from 
the MW/WL database, we find that $\gamma_{\rm mean}$ ($\gamma_{\rm med}$) 
increases from 4{\fdg}4 (5{\fdg}8) to 5{\fdg}8 (6{\fdg}4), 
while $\gamma_{\rm mean}/\vert L\vert_{\rm mean}$ 
($\gamma_{\rm med}/\vert L\vert_{\rm med}$) increases from 0.29 (0.40) to 
0.38 (0.45); likewise, $m_{\rm Joy}^{(\rm mean)}$ ($m_{\rm Joy}^{(\rm med)}$) 
increases from 0.20 (0.27) to 0.32 (0.34).  When the same procedure 
is applied to the DPD/U measurements for cycles~21--23, for which 
33\% of the sunspot groups have $\Delta s < 2{\fdg}5$, 
$\gamma_{\rm mean}$ ($\gamma_{\rm med}$) increases from 5{\fdg}3 (6{\fdg}6) 
to 6{\fdg}4 (7{\fdg}0), $\gamma_{\rm mean}/\vert L\vert_{\rm mean}$ 
($\gamma_{\rm med}/\vert L\vert_{\rm med}$) increases from 0.33 (0.43) 
to 0.40 (0.45), and $m_{\rm Joy}^{(\rm mean)}$ ($m_{\rm Joy}^{(\rm med)}$) 
increases from 0.28 (0.40) to 0.40 (0.45).  The filtering has a smaller 
effect on the tilt angles derived from the DPD/UPP measurements, for which 
only 23\% of the sunspot groups have $\Delta s < 2{\fdg}5$.  In all cases, 
the median tilt angles undergo considerably less change than the 
corresponding means, again suggesting that the median provides a 
more robust statistical measure than the mean.  Figure~9 shows the 
overall effect of omitting sunspot groups with $\Delta s < 2{\fdg}5$ 
on the Joy's law slopes derived from the different databases 
(compare Figure~4).  We note that only 11\% of the MW/MAG measurements 
have the leading- and trailing-polarity centroids separated by 
less than 2{\fdg}5, consistent with the conclusion of Baranyi (2014) 
that most of the MW/WL measurements having $\Delta s < 2{\fdg}5$ 
refer to ``unipolar'' sunspot groups.

Although filtering out the unipolar groups brings the different data sets 
into better agreement with each other, the magnetic measurements still 
appear to provide an upper bound on the tilt angles (see Tables~1 and 2).  
The physical basis of the remaining discrepancy is discussed in Section~4.

\section{CONTRIBUTION OF PLAGE AREAS TO MAGNETIC TILT ANGLES}
Even when new active regions are at their peak development, 
only $\sim$50\% of their magnetic flux is contained in sunspots 
(Schrijver 1987; Sheeley 1966), with the rest being in the form of 
plages/faculae, which are invisible in white-light images except near 
the solar limb.  Plages, best seen in \ion{Ca}{2}~K-line spectroheliograms, 
represent very strong fields that fall below the $\sim$1500~G threshold 
for sunspots.  We now consider the possibility that the magnetic 
tilt angles tend to be larger than their white-light counterparts 
because the magnetograph measurements include the contribution of 
plage areas.  For this purpose, we examine a number of specific cases 
in which magnetic and white-light measurements have been carried out 
for the same active region, but with very different results.  
These examples are all taken from the rising phase of the activity cycle, 
when the tilt angles and the discrepancies between the measurements 
tend to be greatest.

Figure 10 shows a white-light image of active region NOAA 807, 
as recorded by the Debrecen Observatory on 1977 April~18; also displayed 
are longitudinal magnetograms from MW (spatial resolution 3{\farcs}7/pixel) 
and NSO/KP ($\sim$1$^{\prime\prime}$/pixel) taken on the same day.  
The region is centered at latitude $L\simeq -20^\circ$, 
and is dominated by a pair of leading sunspots separated by 4{\fdg}5.  
For this sunspot group, the tilt angles given by the DPD/U 
and DPD/UPP databases are 2{\fdg}5 and 1{\fdg}8, respectively; 
in contrast, Li \& Ulrich (2012) obtained a value of $\gamma$ 
as large as 30$^\circ$ from the MW magnetogram, in agreement with 
the measurement by Sheeley from the KP magnetogram (see Wang \& Sheeley 1989).  
To show why the white-light and magnetic results are so different, 
we have overplotted on each of the images in Figure~10 the axes connecting 
the centroids of the leading and following sectors, according to 
DPD/U, Li \& Ulrich (MW/MAG), and Sheeley (KP/MAG).  It is evident 
that the white-light tilt angles are so small because the easternmost 
of the two sunspots is incorrectly assigned to the following sector.  
From the magnetograms, we see that both spots have leading (negative) 
polarity, while the trailing-polarity sector, centered to the southeast 
of the sunspot pair, consists mainly of plage and is almost invisible 
in the white-light image.

Figure 11 displays a Debrecen white-light image of active region NOAA 882, 
together with the corresponding MW and KP magnetograms.  The region 
is centered at $L\simeq +24^\circ$, and the observations were all made 
on 1977 September~5.  A prominent sunspot is seen only in the 
leading sector.  The tilt angles measured by DPD/U, DPD/UPP, MW/MAG, 
and KP/MAG are 12$^\circ$, 19$^\circ$, 29$^\circ$, and 39$^\circ$, 
respectively.  The DPD/U tilt-angle measurement (not shown) 
can be rejected because it takes the following sector to be centered 
on a small satellite umbra within the leading sunspot, so that 
$\Delta s\sim 0{\fdg}3$.  The DPD/UPP measurement takes into account 
the small pores in the trailing-polarity sector, but not the 
extensive plage areas; the inferred axis (red line in Figure~11) 
is thus less steeply inclined than the axes derived from the MW and 
KP magnetograms (indicated by the blue and green lines, respectively).

Figure 12 shows a Debrecen image of NOAA 1203 on 1978 July 14, 
along with the corresponding MW and KP magnetograms.  For this 
active region, centered at $L\simeq +18^\circ$, the tilt angles 
given by DPD/U, DPD/UPP (not shown), MW/MAG, and KP/MAG are 6$^\circ$, 
5$^\circ$, 21$^\circ$, and 19$^\circ$, respectively.  Here, it is again clear 
that the axial inclination of the BMR cannot be deduced from the 
white-light image in the absence of polarity information.  In this case, 
the large structure dominating the eastern part of the sunspot group 
encloses spots of both polarities, so that the basic assumption 
underlying the white-light measurements---that sunspots of opposite polarity 
do not overlap in longitude---breaks down.  The DPD tilt angles 
are too small because all of the spots located eastward of the 
area-weighted centroid of the sunspot group have automatically been assigned 
to the following-polarity sector.

As an example from the rising phase of cycle 23, Figure 13 shows 
a continuum image and magnetogram of active region NOAA 8086, 
located at $L\simeq +28^\circ$.  Both images were recorded by {\it SOHO}/MDI 
on 1997 September~17.  The sunspot group again consists of a prominent 
leading spot trailed by a smattering of smaller spots/pores.  
The DPD and SDD measurements all give values of $\gamma$ in the range 
$\sim$19$^\circ$--23$^\circ$; the SDD/U tilt angle is indicated by the 
red line connecting the leading spot to the pores located to its northeast.  
In contrast, based on the MDI magnetogram, Li \& Ulrich (2012) derived 
a tilt angle of 41$^\circ$ (indicated by the blue line).  This steeper 
inclination reflects the presence of (positive) leading-polarity plage 
to the east of the large leading spot and (negative) trailing-polarity plage 
to the west of the small trailing spots.

From these case studies, it is clear that the use of white-light 
observations to derive tilt angles is subject to many pitfalls, 
and that the results often do not reflect the actual axial inclinations 
of active regions, as defined by the distribution of magnetic field.  
One of the main sources of error is the well-known asymmetry 
between the leading- and following-polarity sectors, with the latter being 
characterized by smaller spots and less concentrated flux (more extensive 
plage areas) (see Fan et al. 1993, and references therein; 
Murak{\"o}zy et al. 2014).  If only the darkest features (umbrae) are used 
to determine the centroids of the two sectors, as is the case for the 
MW/WL and DPD/U data sets, spots belonging to the leading-polarity sector 
may be assigned to the following sector, particularly when there are 
no visible umbrae of following polarity (as in Figure~10).  Moreover, 
the absence of tilt angle measurements when only a single (or no) umbra 
is observed reduces the size of the sample by as much as $\sim$40\%, 
introducing an unknown bias into the statistics.  Another major source 
of error is associated with active regions where sunspots of 
opposite polarity overlap in longitude, as in Figure~12.  In such cases, 
in the absence of polarity information, the white-light measurements 
will tend to underestimate the actual tilt angles.

As illustrated by the examples of Figures 10, 11, and 13, a more fundamental 
shortcoming of the white-light measurements is that they do not include 
the contribution of plage areas.  That the plage component of 
active regions tends to have a greater axial inclination than the 
sunspot component was pointed out earlier by Howard (1996), 
who compared a histogram of tilt angles derived from MW magnetograms 
taken during 1967--1995 with the tilt angle distribution derived 
from the 1917--1985 MW white-light database (see his Figure~1).  
The daily magnetograms used by Howard were smoothed to a resolution of 
$\gtrsim$3{\fdg}4/pixel, so that sunspots were no longer identifiable 
and active regions effectively consisted entirely of plage fields.  
In general, as illustrated by some of the cases presented here, 
the trailing-polarity sectors of active regions tend to be dominated 
by plages, which occupy a larger area and extend to higher latitudes 
than the sunspots.

It is widely accepted that Joy's law has its physical origin in the 
Coriolis force, which acts to twist toroidally oriented flux tubes 
in the north--south direction as they rise buoyantly to the surface 
(for a dissenting view, see Kosovichev \& Stenflo 2008; Stenflo \& 
Kosovichev 2012).  The tendency for the sunspot component of active regions 
to have a smaller axial inclination than the plage areas may then be 
understood as follows.  As it approaches the surface, an $\Omega$-loop 
expands rapidly in order to maintain pressure equilibrium with 
its surroundings.  The expansion (or diverging flow) in the longitudinal 
direction, with speed $v_{\rm exp}$, gives rise to a Coriolis force 
($\propto v_{\rm exp}$) that acts over a timescale 
$\tau_{\rm exp}\propto v_{\rm exp}^{-1}$ to produce an axial tilt 
$\sin\gamma\propto (v_{\rm exp}\tau_{\rm exp}^2/2)/
(v_{\rm exp}\tau_{\rm exp})\propto v_{\rm exp}^{-1}$.  Since 
$v_{\rm exp}$ (like the buoyant velocity) scales roughly as the 
Alfv{\'e}n speed $v_{\rm A}$, it follows that the tilt angle $\gamma$ 
should scale inversely as the field strength $B$.  Similar results 
are obtained if the limiting effects of magnetic tension and 
drag forces are included (see Fan et al. 1994; Fisher et al. 1995).  
On physical grounds, then, we would expect the tilt angles associated 
with sunspots to be smaller than those associated with the weaker 
plage fields.  It should be emphasized that the axial tilt is here regarded 
to be a function of the local flux density, not of the area-integrated flux.  
Indeed, there is no clear evidence that the tilt angles of BMRs 
vary systematically with their total flux or areal size (Wang \& 
Sheeley 1989, 1991; Kosovichev \& Stenflo 2008; Stenflo \& Kosovichev 2012), 
even though their spread (rms values) increase with decreasing size, 
an effect attributable to random displacements by supergranular convective 
motions (see, e.g., Howard 1989, 1996; Wang \& Sheeley 1989, 1991).

\section{SUMMARY AND DISCUSSION}
Our main objective has been to clarify the differences between 
magnetic and white-light determinations of active-region tilt angles, 
and in particular to understand why magnetograms (used by Wang \& 
Sheeley 1989; Howard 1991a; Tian et al. 2003; Stenflo \& Kosovichev 2012; 
Li \& Ulrich 2012) tend to yield larger inclinations than white-light images 
of sunspot groups (used by Howard 1991b; Dasi-Espuig et al. 2010; 
Ivanov 2012; McClintock \& Norton 2013).  Because the previous 
white-light studies have depended heavily on the historical 
MW and KK databases, we have employed the more recent DPD measurements 
to confirm this systematic effect and to diagnose its causes.  
For comparison with the white-light results, we have adopted the 
tilt angle measurements of Li \& Ulrich (2012) as representative of the 
results of magnetogram-based studies.  We now summarize our conclusions 
(see also Tables~1 and 2).

1. When averaged over cycles 21--23 (1976--2008), the MW magnetic 
measurements yield larger tilt angles and Joy's law slopes than the 
DPD white-light observations.  The mean (median) tilt angle ranges from 
6{\fdg}2$\pm$0{\fdg}2 (7{\fdg}5$\pm$0{\fdg}2) for MW/MAG, to 
5{\fdg}9$\pm$0{\fdg}1 (6{\fdg}9$\pm$0{\fdg}2) for DPD/UPP, to 
5{\fdg}3$\pm$0{\fdg}2 (6{\fdg}6$\pm$0{\fdg}2) for DPD/U.  
The mean-based (median-based) values of $m_{\rm Joy}$ are found to be 
0.46$\pm$0.03 (0.52$\pm$0.03) for MW/MAG, 0.32$\pm 0.05$ (0.40$\pm$0.03) 
for DPD/UPP, and 0.28$\pm$0.03 (0.40$\pm$0.04) for DPD/U.

2. The historical MW white-light database gives systematically smaller 
tilt angles and Joy's law slopes than the DPD measurements.  When averaged 
over cycles~16--21 (1923--1985), MW/WL yields a mean (median) tilt angle 
of 4{\fdg}4$\pm$0{\fdg}2 (5{\fdg}8$\pm$0{\fdg}2) and a 
mean-based (median-based) Joy's law slope of 0.20$\pm$0.01 (0.27$\pm$0.01).  
However, the MW and DPD data sets show surprisingly good agreement 
where they overlap in cycle~21 (see Figures~4--6).

3. In 43\% of the MW sunspot-group measurements for cycles 16--21,
the angular separation between the leading and following parts 
is less than 2{\fdg}5.  In the great majority ($\sim$84\%) of these cases, 
as shown by Baranyi (2014), the two sectors are likely to have 
the same polarity.  Filtering out all regions with $\Delta s < 2{\fdg}5$ 
from the MW/WL and DPD/U data sets substantially increases 
both the average tilt angles and the Joy's law slopes.  
That a large fraction of the umbral-based measurements refer to 
``unipolar'' groups reflects the tendency for well-defined sunspots to be 
found preferentially in the leading-polarity sectors of active regions.

4. In view of the skewed nature of the measured distributions, 
which are centered on positive values of $\gamma$ and are thus asymmetric 
with respect to the imposed endpoints at $\gamma = \pm 90^\circ$, 
medians provide a better characterization of the tilt angle properties 
than simple arithmetic means (as was recognized earlier by 
Li \& Ulrich 2012).  As evident from a comparison of Tables~1 and 2, 
the median-based values of $m_{\rm Joy}$, $\gamma$, and 
$\gamma/\vert L\vert$ tend to be substantially larger than the corresponding 
mean-based values.  It is also noteworthy that the medians change less 
than the means when sunspot groups with $\Delta s < 2{\fdg}5$ are omitted.

5. That the tilt angles derived from magnetograms remain somewhat larger 
than those deduced from white-light images, even after filtering out 
unipolar sunspot groups, is due to the contribution of plages, 
which are seen in magnetograms but not in white light.  The plage areas 
of an active region tend to have a larger axial inclination 
than the sunspots.  This property was first noted by Howard (1996), 
although he did not correct the white-light tilt angles for the 
contribution of unipolar groups.

6. The difference between the sunspot and plage tilt angles can be 
attributed to the action of the Coriolis force, which imparts to 
each buoyant flux tube a twist proportional to its rise/expansion time.  
Since $\tau_{\rm exp}$ scales inversely as the local field strength, 
the weaker, more slowly rising/expanding fields that form plages 
end up with a greater net twist than the stronger sunspot fields.  
The same holds if the twist due to the Coriolis force is limited 
by magnetic tension (see, e.g., Fan et al. 1994).

7. Given the large uncertainties involved in measuring tilt angles, 
whether white-light images or magnetograms are employed, it remains unclear 
whether Joy's law undergoes physically significant variations 
between different cycles.  However, Figure~6 suggests that the 
cycle-averaged, latitude-normalized parameters 
$\gamma_{\rm mean}/\vert L\vert_{\rm mean}$ ($\sim$0.3--0.4) and 
$\gamma_{\rm med}/\vert L\vert_{\rm med}$ ($\sim$0.4--0.5) remain 
relatively constant from cycle to cycle.

Our main conclusion is that white-light measurements generally underestimate 
the axial inclinations of active regions, even when the leading- 
and following-polarity sectors are correctly identified, 
because the sunspot component is characterized by a smaller tilt 
than the plage component.  By the same token, measurements employing 
magnetograms that underestimate the flux in sunspots (due to 
low spatial resolution or saturation effects) may slightly overestimate 
the axial tilts of active regions.

It is apparent that the tilt angles extracted from the 
MW white-light database do not provide a reliable basis for 
long-term flux-transport simulations, or for deducing cycle-to-cycle 
relationships among tilt angles, cycle amplitudes/total sunspot areas, 
and the polar fields (see Dasi-Espuig et al. 2010; 
Mu{\~n}oz-Jaramillo et al. 2013).  However, these measurements 
could be improved by exploiting the polarity information available 
in the daily MW sunspot-group drawings.  In addition, use should be made 
of the digitized MW \ion{Ca}{2} K-line images spanning the period 
1915--1985,\footnote{See \texttt{http://ulrich.astro.ucla.edu/MW\_SPADP}.} 
which show the plage areas associated with the sunspot groups, 
while at the same time outlining more clearly the boundaries of the 
active regions (see, e.g., Figures~1 to 12 in Sheeley et al. 2011).  
It would be particularly instructive to compare the tilt angles 
defined by the \ion{Ca}{2} emission contours with those derived from 
the white-light sunspots.

From an analysis of white-light facular data compiled by the 
Royal Greenwich Observatory, Foukal (1993) concluded that the ratio 
of facular to sunspot area decreased during the highest-amplitude cycles 
(18 and 19).  Since faculae are the white-light counterparts (visible near 
the limb) of plages and strong magnetic network, such a decrease would 
imply that the axial tilts of active regions should be somewhat smaller 
during very active cycles than in weaker cycles.  However, recent 
photometric measurements by Chapman et al. (2011) indicate that 
the facular/plage area in fact increases linearly with sunspot area, 
without flattening out for large total sunspot areas.

An important question not addressed in this study is the effect 
of evolutionary processes on the measured tilt angles.  That the values 
obtained depend on when the measurements are made was already 
emphasized by Brunner (1930), who noted that the axial inclinations of 
emerging sunspot groups decrease with time due to the westward proper motions 
of their leading spots (see also Kosovichev \& Stenflo 2008).  
The slope derived for Joy's law depends sensitively on the relatively 
few active regions at latitudes $\vert L\vert\gtrsim 25^\circ$, 
which are also those most strongly affected by differential rotation.  
For a BMR centered at $\vert L\vert = 27{\fdg}5$ with a 
latitudinal pole separation of 2$^\circ$, for example, $\Delta\phi$ increases 
at a rate of $\sim$0{\fdg}09~day$^{-1}$, so that rotational shearing 
may have a noticeable effect on the inclinations of regions older than 
$\sim$10~days.  Tilt angle measurements should also be performed before 
supergranular diffusion has converted the sunspot fields into network/plage.

It is clear from this study that reliable and physically meaningful 
tilt angles cannot be derived from white-light images alone, 
which not only lack the essential polarity information, 
but also omit the plage component of active regions.  Even when 
magnetograms are employed, care is needed when applying automated techniques.  
Ideally, each active region should be tracked from day to day 
and its axial inclination measured at the time of peak development, 
after the flux has fully emerged but before it begins to decay through 
transport processes.  Such tracking, which also makes it easier 
to identify the boundaries of closely spaced active regions, is probably 
best done by human eye; a computer algorithm can then be employed 
to determine the centroids of the leading- and trailing-polarity flux 
within the visually checked/adjusted active-region boundaries.

It will be especially interesting to compare the magnetic tilt angles 
determined for the current weak cycle with those derived for 
cycles~21--23.  Accurate measurements using a combination of visual 
and automated techniques may help to resolve the question of 
whether cycle-to-cycle variations in the polar fields are 
related to systematic variations in the axial tilts of active regions.

\acknowledgments
We thank the referee for comments, and G. Chintzoglou, J. S. Morrill, 
K. Muglach, N. R. Sheeley, Jr., D. G. Socker, R. K. Ulrich, and the 
participants of the International Space Science Institute Workshop 
on ``The Solar Activity Cycle: Physical Causes and Consequences'' 
for helpful discussions.  We are also indebted to the 
Debrecen Heliophysical Observatory, MW/UCLA, and NSO/KP 
for providing the data used in this investigation, which was funded 
by the Office of Naval Research and by the European Community's 
Seventh Framework Programme (project eHEROES).

\newpage
\begin{deluxetable}{cccccccccccc}
\rotate
\tabletypesize{\scriptsize}
\tablewidth{0pt}
\tablecaption{Tilt Angle Parameters (Mean-Based)}
\tablehead{
\colhead{} & \colhead{} & \multicolumn{10}{c}{Cycle Number} \\
\cline{3-12} \\
\colhead{Parameter} & \colhead{Database} &
\colhead{16} & \colhead{17} & \colhead{18} & \colhead{19} & \colhead{20} & 
\colhead{21} & \colhead{22} & \colhead{23} & 16--21 & 21--23}
\startdata

$m_{\rm Joy}^{(\rm mean)}$ & MW/MAG & \nodata & \nodata & \nodata & 
\nodata & \nodata & 0.48$\pm$0.05 & 0.37$\pm$0.07 & 0.50$\pm$0.06 & 
\nodata & 0.46$\pm$0.03 \\

  & ($\Delta s > 2{\fdg}5$) & \nodata & \nodata & \nodata & 
\nodata & \nodata & 0.47$\pm$0.03 & 0.42$\pm$0.06 & 0.53$\pm$0.07 & 
\nodata & 0.48$\pm$0.04 \\

  & MW/WL & 0.28$\pm$0.11 & 0.17$\pm$0.05 & 0.21$\pm$0.08 & 
0.19$\pm$0.02 & 0.14$\pm$0.04 & 0.27$\pm$0.06 & \nodata & \nodata & 
0.20$\pm$0.01 & \nodata \\

  & ($\Delta s > 2{\fdg}5$) & 0.51$\pm$0.12 & 0.43$\pm$0.08 & 
0.20$\pm$0.02 & 0.27$\pm$0.06 & 0.36$\pm$0.04 & 0.34$\pm$0.08 & 
\nodata & \nodata & 0.32$\pm$0.04 & \nodata \\

  & DPD/U & \nodata & \nodata & \nodata & \nodata & \nodata & 
0.16$\pm$0.06 & 0.36$\pm$0.05 & 0.33$\pm$0.08 & \nodata & 0.28$\pm$0.03 \\

  & ($\Delta s > 2{\fdg}5$) & \nodata & \nodata & \nodata & \nodata & 
\nodata & 0.31$\pm$0.02 & 0.47$\pm$0.05 & 0.43$\pm$0.05 & \nodata & 
0.40$\pm$0.03 \\

  & DPD/UPP & \nodata & \nodata & \nodata & \nodata & \nodata & 
0.27$\pm$0.07 & 0.32$\pm$0.04 & 0.38$\pm$0.05 & \nodata & 0.32$\pm$0.05 \\

  & ($\Delta s > 2{\fdg}5$) & \nodata & \nodata & \nodata & 
\nodata & \nodata & 0.28$\pm$0.05 & 0.37$\pm$0.04 & 0.41$\pm$0.03 & 
\nodata & 0.35$\pm$0.04 \\[0.15cm]

$\gamma_{\rm mean}$ & MW/MAG & \nodata & \nodata & \nodata & \nodata & 
\nodata & 5{\fdg}8$\pm$0{\fdg}3 & 6{\fdg}8$\pm$0{\fdg}3 & 
6{\fdg}1$\pm$0{\fdg}3 & \nodata & 6{\fdg}2$\pm$0{\fdg}2 \\

  & ($\Delta s > 2{\fdg}5$) & \nodata & \nodata & \nodata & 
\nodata & \nodata & 6{\fdg}0$\pm$0{\fdg}3 & 7{\fdg}0$\pm$0{\fdg}3 & 
6{\fdg}3$\pm$0{\fdg}3 & \nodata & 6{\fdg}4$\pm$0{\fdg}2 \\

  & MW/WL & 4{\fdg}4$\pm$0{\fdg}6 & 4{\fdg}7$\pm$0{\fdg}5 & 
4{\fdg}8$\pm$0{\fdg}5 & 3{\fdg}6$\pm$0{\fdg}4 & 4{\fdg}1$\pm$0{\fdg}5 & 
5{\fdg}3$\pm$0{\fdg}5 & \nodata & \nodata & 4{\fdg}4$\pm$0{\fdg}2 & 
\nodata \\

  & ($\Delta s > 2{\fdg}5$) & 6{\fdg}6$\pm$0{\fdg}5 & 
5{\fdg}5$\pm$0{\fdg}5 & 6{\fdg}4$\pm$0{\fdg}4 & 4{\fdg}6$\pm$0{\fdg}4 & 
5{\fdg}5$\pm$0{\fdg}4 & 6{\fdg}4$\pm$0{\fdg}5 & \nodata & \nodata & 
5{\fdg}8$\pm$0{\fdg}2 & \nodata \\

  & DPD/U & \nodata & \nodata & \nodata & \nodata & \nodata & 
5{\fdg}2$\pm$0{\fdg}3 & 5{\fdg}1$\pm$0{\fdg}3 & 5{\fdg}6$\pm$0{\fdg}3 & 
\nodata & 5{\fdg}3$\pm$0{\fdg}2 \\

  & ($\Delta s > 2{\fdg}5$) & \nodata & \nodata & \nodata & \nodata & 
\nodata & 6{\fdg}6$\pm$0{\fdg}3 & 6{\fdg}2$\pm$0{\fdg}2 & 
6{\fdg}4$\pm$0{\fdg}2 & \nodata & 6{\fdg}4$\pm$0{\fdg}1 \\

  & DPD/UPP & \nodata & \nodata & \nodata & \nodata & \nodata & 
5{\fdg}5$\pm$0{\fdg}2 & 6{\fdg}1$\pm$0{\fdg}2 & 6{\fdg}3$\pm$0{\fdg}2 & 
\nodata & 5{\fdg}9$\pm$0{\fdg}1 \\

  & ($\Delta s > 2{\fdg}5$) & \nodata & \nodata & \nodata & 
\nodata & \nodata & 6{\fdg}1$\pm$0{\fdg}2 & 6{\fdg}7$\pm$0{\fdg}2 & 
7{\fdg}0$\pm$0{\fdg}2 & \nodata & 6{\fdg}6$\pm$0{\fdg}1 \\[0.15cm]

$\gamma_{\rm mean}/\vert L\vert_{\rm mean}$ & MW/MAG & \nodata & \nodata & 
\nodata & \nodata & \nodata & 0.37$\pm$0.02 & 0.41$\pm$0.02 & 
0.39$\pm$0.02 & \nodata & 0.39$\pm$0.01 \\

  & ($\Delta s > 2{\fdg}5$) & \nodata & \nodata & \nodata & 
\nodata & \nodata & 0.38$\pm$0.02 & 0.42$\pm$0.02 & 0.40$\pm$0.02 & 
\nodata & 0.40$\pm$0.01 \\

  & MW/WL & 0.31$\pm$0.04 & 0.32$\pm$0.04 & 0.32$\pm$0.03 & 
0.22$\pm$0.02 & 0.29$\pm$0.03 & 0.35$\pm$0.03 & \nodata & \nodata & 
0.29$\pm$0.01 & \nodata \\

  & ($\Delta s > 2{\fdg}5$) & 0.46$\pm$0.04 & 0.38$\pm$0.03 & 
0.43$\pm$0.03 & 0.28$\pm$0.02 & 0.40$\pm$0.03 & 0.43$\pm$0.03 & 
\nodata & \nodata & 0.38$\pm$0.01 & \nodata \\

  & DPD/U & \nodata & \nodata & \nodata & \nodata & \nodata & 
0.33$\pm$0.02 & 0.31$\pm$0.02 & 0.36$\pm$0.02 & \nodata & 0.33$\pm$0.01 \\

  & ($\Delta s > 2{\fdg}5$) & \nodata & \nodata & \nodata & \nodata & 
\nodata & 0.42$\pm$0.02 & 0.37$\pm$0.01 & 0.40$\pm$0.02 & \nodata & 
0.40$\pm$0.01 \\

  & DPD/UPP & \nodata & \nodata & \nodata & \nodata & \nodata & 
0.35$\pm$0.01 & 0.37$\pm$0.01 & 0.40$\pm$0.01 & \nodata & 0.37$\pm$0.01 \\

  & ($\Delta s > 2{\fdg}5$) & \nodata & \nodata & \nodata & 
\nodata & \nodata & 0.39$\pm$0.01 & 0.41$\pm$0.01 & 0.45$\pm$0.01 & 
\nodata & 0.42$\pm$0.01 \\

\enddata
\tablecomments{Here and throughout this study, only measurements 
made within 45$^\circ$ of central meridian longitude are included 
in the statistics.  Quoted errors, based on the number of measurements 
and their scatter, are 1$\sigma$; the actual uncertainties are probably 
much larger.  The MW/WL and DPD/U tilt angles are derived from the 
area-weighted positions of umbrae only, while the DPD/UPP measurements 
include umbrae, penumbrae, and faint pores.  The second row of values 
provided for each database shows the result of omitting all regions 
whose leading and following centroids are separated by less than 2{\fdg}5 
(see text).  Cycle dates are: 1923~August--1933~August (16), 
1933~September--1944~January (17), 1944~February--1954~March (18), 
1954~April--1964~September (19), 1964~October--1976~May (20), 
1976~June--1986~August (21), 1986~September--1996~April (22), 
1996~May--2008~December (23).}
\end{deluxetable}

\newpage
\begin{deluxetable}{cccccccccccc}
\rotate
\tabletypesize{\scriptsize}
\tablewidth{0pt}
\tablecaption{Tilt Angle Parameters (Median-Based)}
\tablehead{
\colhead{} & \colhead{} & \multicolumn{10}{c}{Cycle Number} \\
\cline{3-12} \\
\colhead{Parameter} & \colhead{Database} & 
\colhead{16} & \colhead{17} & \colhead{18} & \colhead{19} & \colhead{20} & 
\colhead{21} & \colhead{22} & \colhead{23} & 16--21 & 21--23}
\startdata

$m_{\rm Joy}^{(\rm med)}$ & MW/MAG & \nodata & \nodata & \nodata & 
\nodata & \nodata & 0.57$\pm$0.03 & 0.47$\pm$0.05 & 0.48$\pm$0.05 & 
\nodata & 0.52$\pm$0.03 \\

  & ($\Delta s > 2{\fdg}5$) & \nodata & \nodata & \nodata & 
\nodata & \nodata & 0.57$\pm$0.03 & 0.51$\pm$0.04 & 0.49$\pm$0.05 & 
\nodata & 0.53$\pm$0.03 \\

  & MW/WL & 0.41$\pm$0.12 & 0.30$\pm$0.02 & 0.14$\pm$0.09 & 
0.24$\pm$0.05 & 0.21$\pm$0.07 & 0.41$\pm$0.06 & \nodata & \nodata & 
0.27$\pm$0.01 & \nodata \\

  & ($\Delta s > 2{\fdg}5$) & 0.47$\pm$0.14 & 0.46$\pm$0.08 & 
0.16$\pm$0.02 & 0.29$\pm$0.07 & 0.37$\pm$0.06 & 0.45$\pm$0.08 & 
\nodata & \nodata & 0.34$\pm$0.04 & \nodata \\

  & DPD/U & \nodata & \nodata & \nodata & \nodata & \nodata & 
0.34$\pm$0.02 & 0.46$\pm$0.06 & 0.41$\pm$0.06 & \nodata & 0.40$\pm$0.04 \\

  & ($\Delta s > 2{\fdg}5$) & \nodata & \nodata & \nodata & \nodata & 
\nodata & 0.41$\pm$0.04 & 0.50$\pm$0.06 & 0.46$\pm$0.04 & \nodata & 
0.45$\pm$0.04 \\

  & DPD/UPP & \nodata & \nodata & \nodata & \nodata & \nodata & 
0.36$\pm$0.03 & 0.37$\pm$0.05 & 0.44$\pm$0.04 & \nodata & 0.40$\pm$0.03 \\

  & ($\Delta s > 2{\fdg}5$) & \nodata & \nodata & \nodata & 
\nodata & \nodata & 0.38$\pm$0.04 & 0.35$\pm$0.04 & 0.44$\pm$0.02 & 
\nodata & 0.40$\pm$0.03 \\[0.15cm]

$\gamma_{\rm med}$ & MW/MAG & \nodata & \nodata & \nodata & \nodata & 
\nodata & 7{\fdg}5$\pm$0{\fdg}4 & 7{\fdg}8$\pm$0{\fdg}4 & 
7{\fdg}1$\pm$0{\fdg}3 & \nodata & 7{\fdg}5$\pm$0{\fdg}2 \\

  & ($\Delta s > 2{\fdg}5$) & \nodata & \nodata & \nodata & 
\nodata & \nodata & 7{\fdg}55$\pm$0{\fdg}4 & 7{\fdg}8$\pm$0{\fdg}4 & 
7{\fdg}1$\pm$0{\fdg}3 & \nodata & 7{\fdg}5$\pm$0{\fdg}2 \\

  & MW/WL & 6{\fdg}1$\pm$0{\fdg}7 & 5{\fdg}3$\pm$0{\fdg}7 & 
6{\fdg}4$\pm$0{\fdg}6 & 4{\fdg}7$\pm$0{\fdg}5 & 5{\fdg}5$\pm$0{\fdg}6 & 
7{\fdg}4$\pm$0{\fdg}6 & \nodata & \nodata & 5{\fdg}8$\pm$0{\fdg}2 & \nodata \\

  & ($\Delta s > 2{\fdg}5$) & 7{\fdg}4$\pm$0{\fdg}7 & 
5{\fdg}85$\pm$0{\fdg}6 & 7{\fdg}4$\pm$0{\fdg}5 & 5{\fdg}1$\pm$0{\fdg}5 & 
5{\fdg}8$\pm$0{\fdg}5 & 7{\fdg}8$\pm$0{\fdg}6 & \nodata & \nodata & 
6{\fdg}4$\pm$0{\fdg}2 & \nodata \\

  & DPD/U & \nodata & \nodata & \nodata & \nodata & \nodata & 
6{\fdg}7$\pm$0{\fdg}4 & 6{\fdg}0$\pm$0{\fdg}4 & 7{\fdg}0$\pm$0{\fdg}4 & 
\nodata & 6{\fdg}6$\pm$0{\fdg}2 \\

  & ($\Delta s > 2{\fdg}5$) & \nodata & \nodata & \nodata & \nodata & 
\nodata & 7{\fdg}2$\pm$0{\fdg}3 & 6{\fdg}4$\pm$0{\fdg}3 & 
7{\fdg}25$\pm$0{\fdg}3 & \nodata & 7{\fdg}0$\pm$0{\fdg}2 \\

  & DPD/UPP & \nodata & \nodata & \nodata & \nodata & \nodata & 
6{\fdg}8$\pm$0{\fdg}3 & 6{\fdg}5$\pm$0{\fdg}3 & 7{\fdg}4$\pm$0{\fdg}3 & 
\nodata & 6{\fdg}9$\pm$0{\fdg}2 \\

  & ($\Delta s > 2{\fdg}5$) & \nodata & \nodata & \nodata & 
\nodata & \nodata & 7{\fdg}2$\pm$0{\fdg}3 & 6{\fdg}9$\pm$0{\fdg}3 & 
7{\fdg}8$\pm$0{\fdg}3 & \nodata & 7{\fdg}3$\pm$0{\fdg}2 \\[0.15cm]

$\gamma_{\rm med}/\vert L\vert_{\rm med}$ & MW/MAG & \nodata & \nodata & 
\nodata & \nodata & \nodata & 0.50$\pm$0.02 & 0.48$\pm$0.02 & 
0.48$\pm$0.02 & \nodata & 0.48$\pm$0.01 \\

  & ($\Delta s > 2{\fdg}5$) & \nodata & \nodata & \nodata & 
\nodata & \nodata & 0.49$\pm$0.02 & 0.48$\pm$0.02 & 0.48$\pm$0.02 & 
\nodata & 0.48$\pm$0.01 \\

  & MW/WL & 0.44$\pm$0.05 & 0.38$\pm$0.05 & 0.44$\pm$0.04 & 
0.30$\pm$0.03 & 0.41$\pm$0.04 & 0.51$\pm$0.04 & \nodata & \nodata & 
0.40$\pm$0.02 & \nodata \\

  & ($\Delta s > 2{\fdg}5$) & 0.53$\pm$0.05 & 0.43$\pm$0.04 & 
0.53$\pm$0.04 & 0.32$\pm$0.03 & 0.44$\pm$0.04 & 0.54$\pm$0.04 & 
\nodata & \nodata & 0.45$\pm$0.02 & \nodata \\

  & DPD/U & \nodata & \nodata & \nodata & \nodata & \nodata & 
0.45$\pm$0.03 & 0.38$\pm$0.02 & 0.47$\pm$0.02 & \nodata & 0.43$\pm$0.01 \\

  & ($\Delta s > 2{\fdg}5$) & \nodata & \nodata & \nodata & \nodata & 
\nodata & 0.48$\pm$0.02 & 0.40$\pm$0.02 & 0.47$\pm$0.02 & \nodata & 
0.45$\pm$0.01 \\

  & DPD/UPP & \nodata & \nodata & \nodata & \nodata & \nodata & 
0.45$\pm$0.02 & 0.41$\pm$0.02 & 0.49$\pm$0.02 & \nodata & 0.45$\pm$0.01 \\

  & ($\Delta s > 2{\fdg}5$) & \nodata & \nodata & \nodata & 
\nodata & \nodata & 0.48$\pm$0.02 & 0.44$\pm$0.02 & 0.51$\pm$0.02 & 
\nodata & 0.48$\pm$0.01 \\

\enddata
\end{deluxetable}

\newpage
\begin{figure}
\vspace{-3.5cm}
\centerline{\includegraphics[width=40pc]{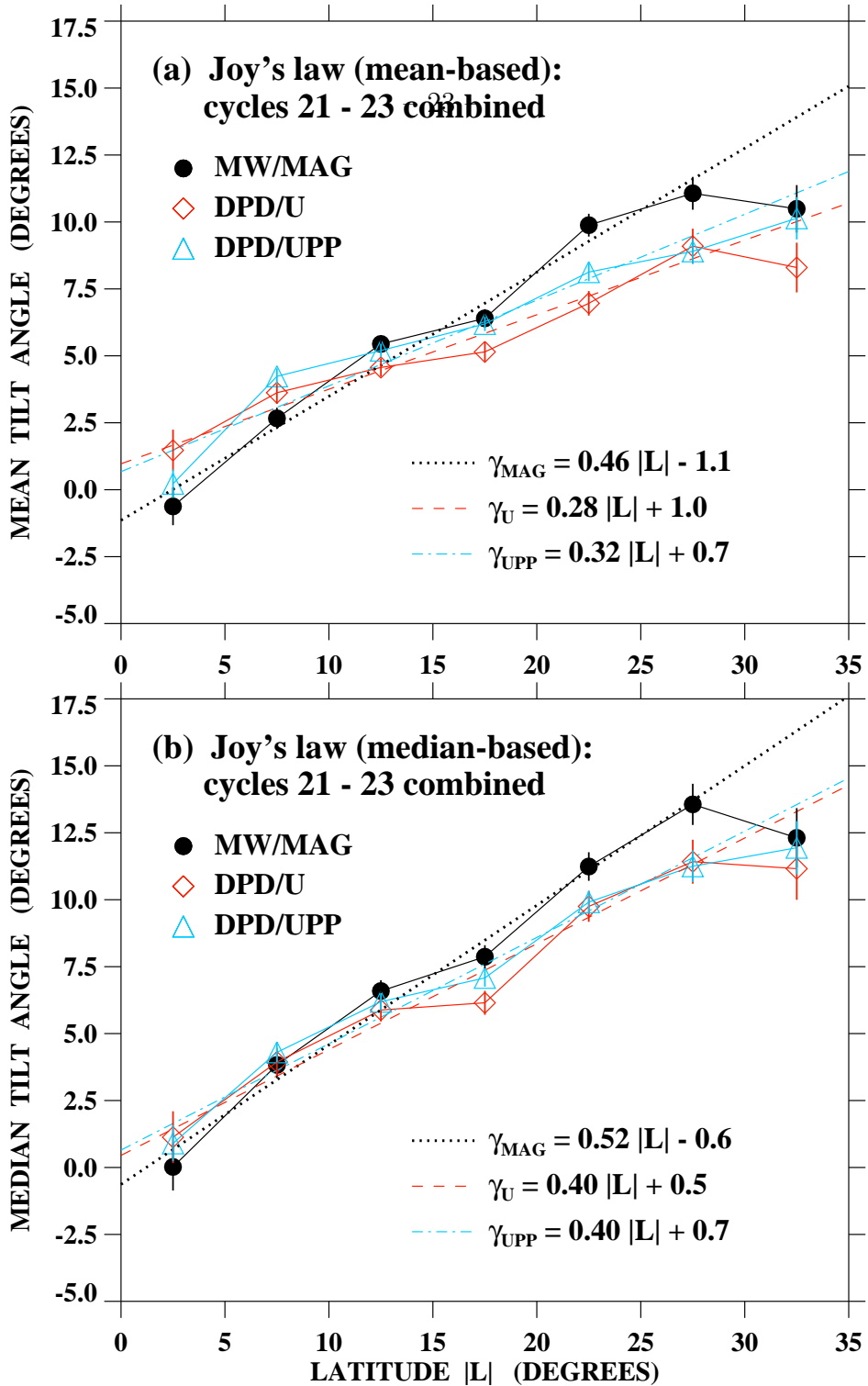}}
\vspace{-1.0cm}
\caption{Joy's law for cycles 21--23 combined, derived from the 
MW magnetic measurements of Li \& Ulrich (2012: MW/MAG), and from 
Debrecen Observatory white-light measurements based on dark umbrae only 
(DPD/U) and on umbrae, penumbrae, and pores (DPD/UPP).  
(a) Mean-based Joy's law: arithmetic means of the tilt angle measurements 
have been taken over 5$^\circ$-wide bins in unsigned latitude $\vert L\vert$, 
and a least-squares linear fit applied to the six points spanning the range 
0$^\circ$--30$^\circ$.  Vertical bars indicate one standard error of 
the mean.  (b) Median-based Joy's law: here, median values of the 
measured tilt angles have been calculated for each latitude bin 
and a least-squares fit applied over the range 0$^\circ$--30$^\circ$.  
Vertical bars represent one standard error of the median.}
\end{figure}

\newpage
\begin{figure}
\vspace{-4.8cm}
\centerline{\includegraphics[width=40pc]{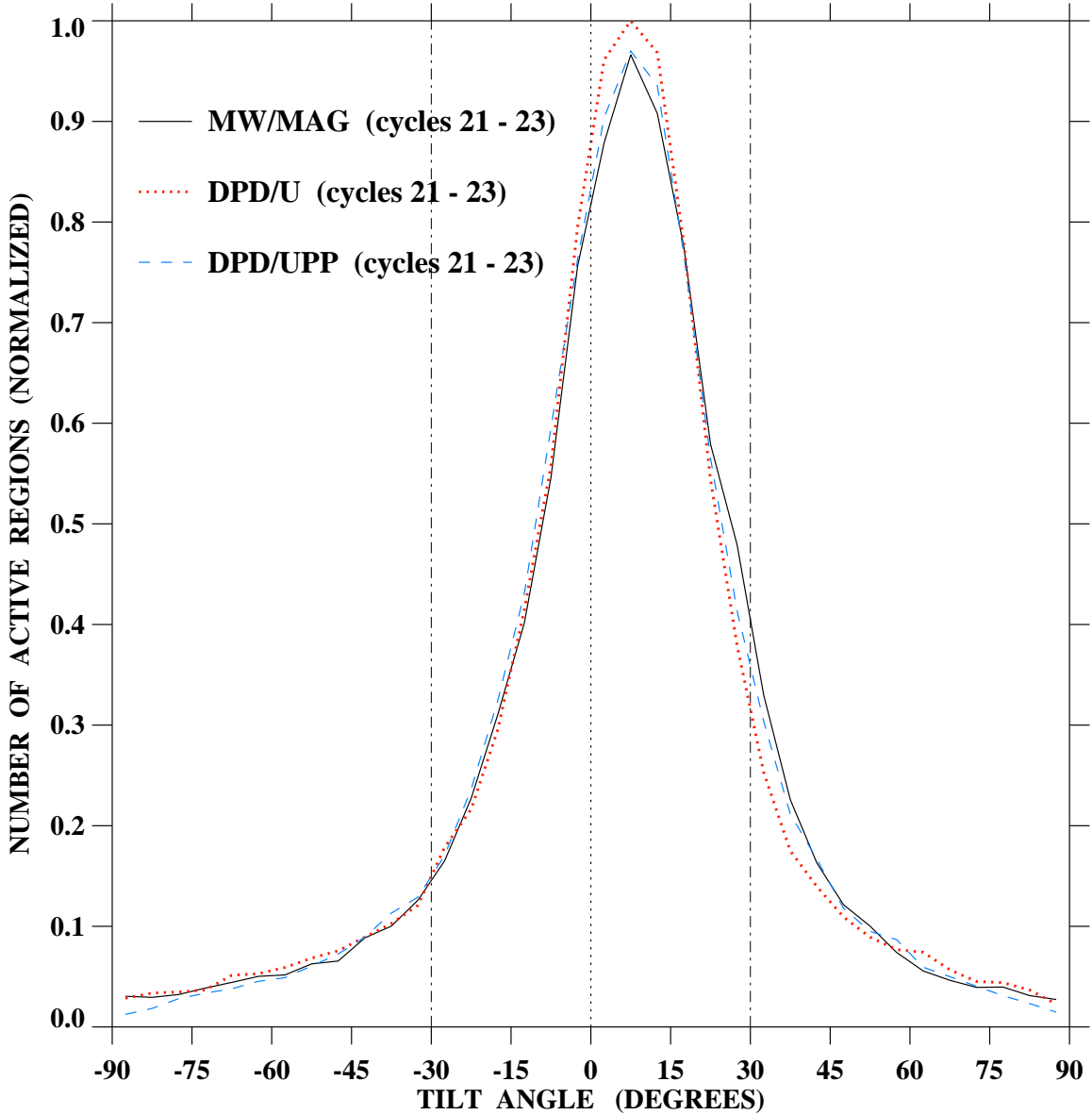}}
\vspace{-2.2cm}
\caption{Histograms (frequency distributions) of active-region tilt angles 
measured during 1976--2008 (cycles~21--23), normalized by dividing by 
the total number of measurements in each data set and multiplying by 
an arbitrary constant.  Black solid curve: MW/MAG (HWHM = 17{\fdg}5).  
Red dotted curve: DPD/U (HWHM = 15{\fdg}4).  Blue dashed curve: 
DPD/UPP (HWHM = 16{\fdg}9).  Bin size for the histograms is 5$^\circ$ 
(the discrete bin points are connected by a smooth curve rather than 
being plotted in steplike fashion).  Vertical dotted (dash-dotted) lines 
mark the location of $\gamma = 0^\circ$ ($\gamma = \pm 30^\circ$).  
Significant differences between the three Gaussian-like distributions 
occur over the interval $+25^\circ\lesssim\gamma\lesssim +40^\circ$, 
with DPD/UPP and especially MW/MAG showing a greater relative number 
of these large tilt angles than DPD/U, which makes up for its deficit 
with a surplus at small tilt angles.  The restriction of $\gamma$ 
to the range between $-90^\circ$ and $+90^\circ$ causes the 
distribution functions to be skewed leftward relative to their peaks, 
which in turn explains why their means are smaller than their medians.}
\end{figure}

\newpage
\begin{figure}
\vspace{-2.2cm}
\centerline{\includegraphics[width=40pc]{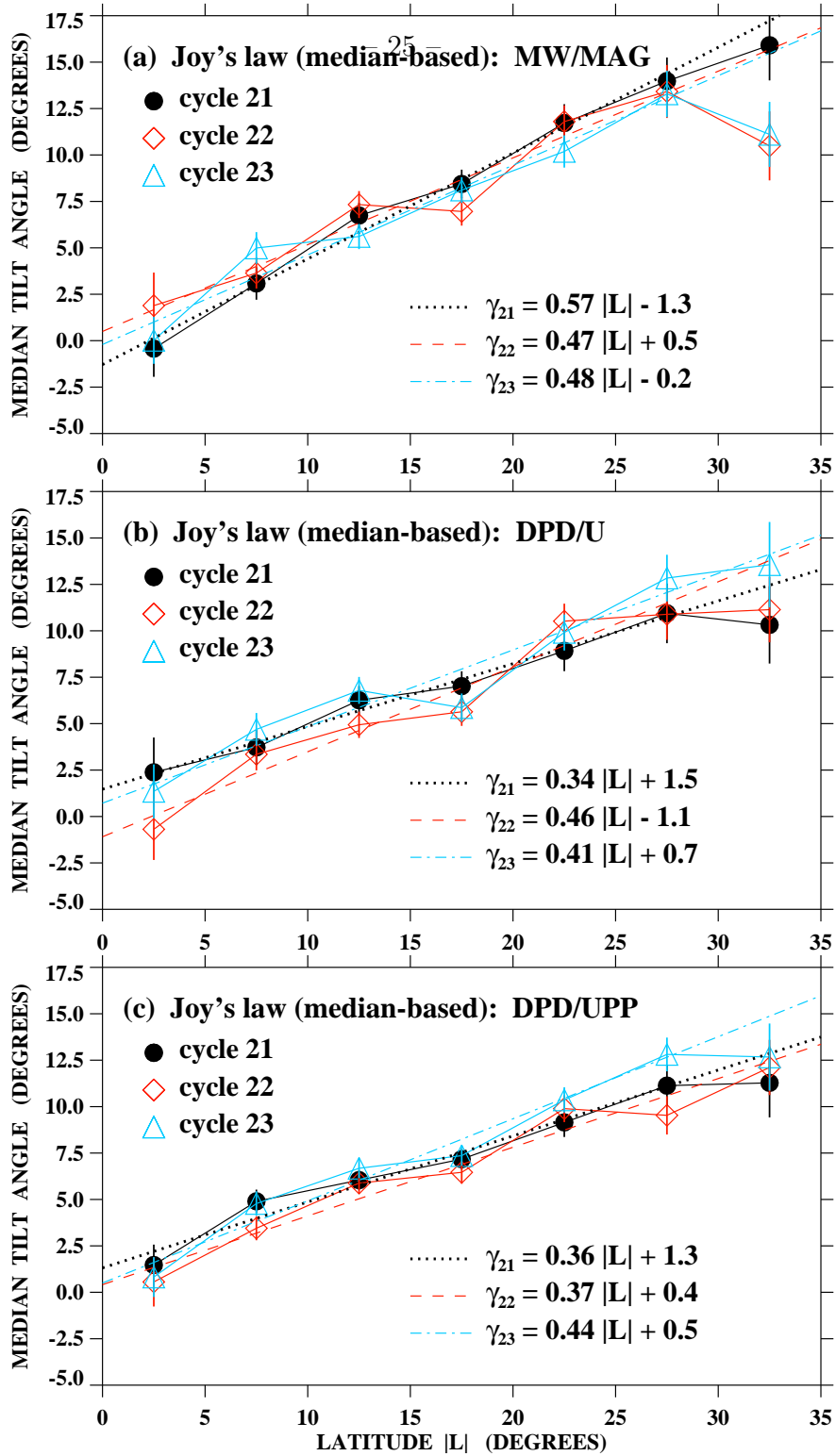}}
\vspace{-1.0cm}
\caption{Median-based Joy's law derived separately for cycle 21 
(1976 June--1986 August), cycle~22 (1986~September--1996~April), and 
cycle~23 (1996~May--2008~December).  (a) MW/MAG measurements 
(Li \& Ulrich 2012).  (b) DPD/U white-light measurements (including 
dark umbrae only).  (c) DPD/UPP white-light measurements (including 
umbrae, penumbrae, and faint pores).}
\end{figure}

\newpage
\begin{figure}
\vspace{-2.0cm}
\centerline{\includegraphics[width=40pc]{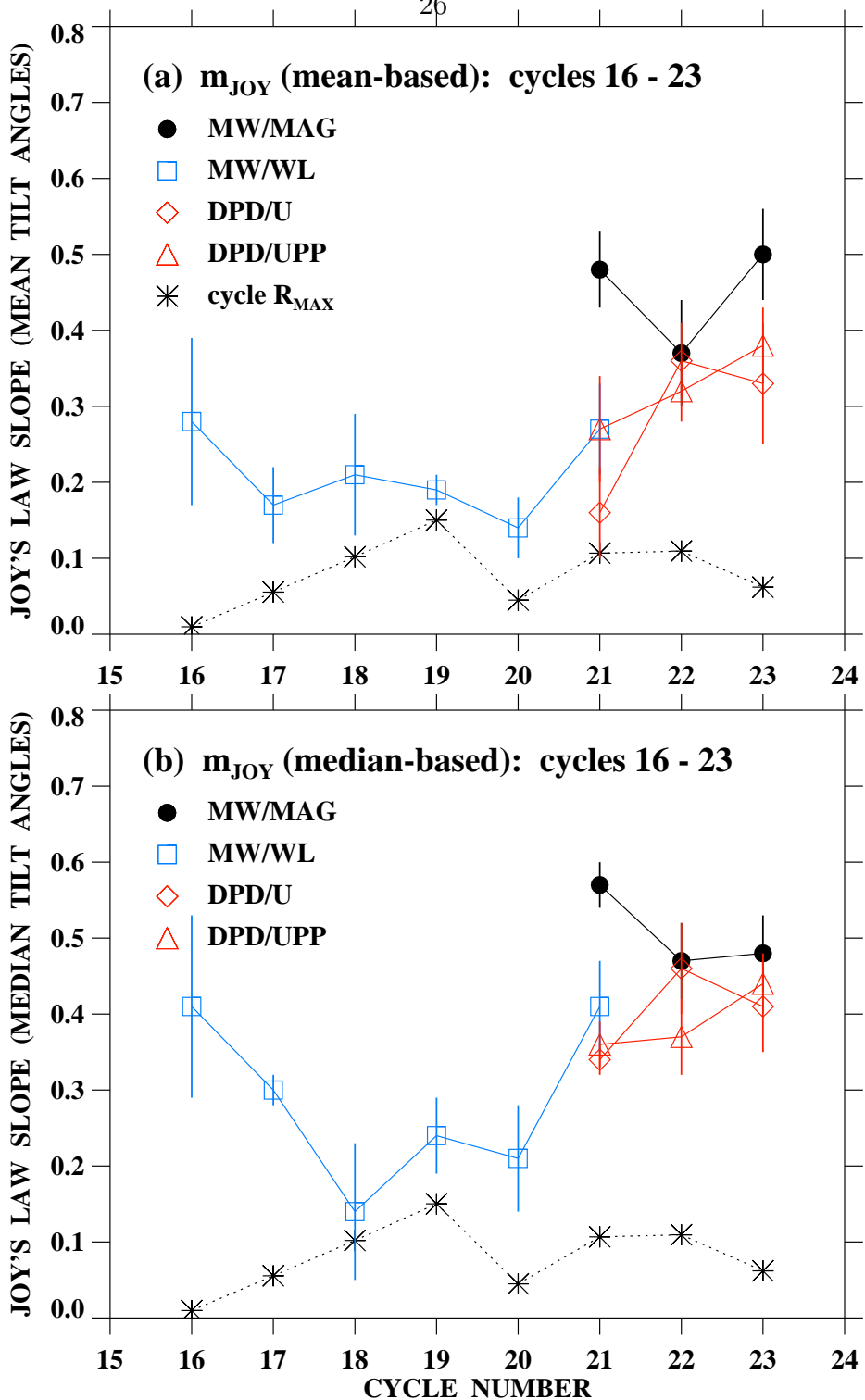}}
\vspace{-1.0cm}
\caption{Cycle-averaged Joy's law slopes, derived from MW/MAG, MW/WL, 
DPD/U, and DPD/UPP measurements.  (a) Mean-based values 
$m_{\rm Joy}^{(\rm mean)}$.  (b) Median-based values 
$m_{\rm Joy}^{(\rm med)}$.  Vertical bars show 1$\sigma$ errors.  
Asterisks indicate the (arbitrarily shifted and scaled) 
maximum yearly sunspot numbers $R_{\rm max}$ for cycles~16--23.}
\end{figure}

\newpage
\begin{figure}
\vspace{-2.5cm}
\centerline{\includegraphics[width=40pc]{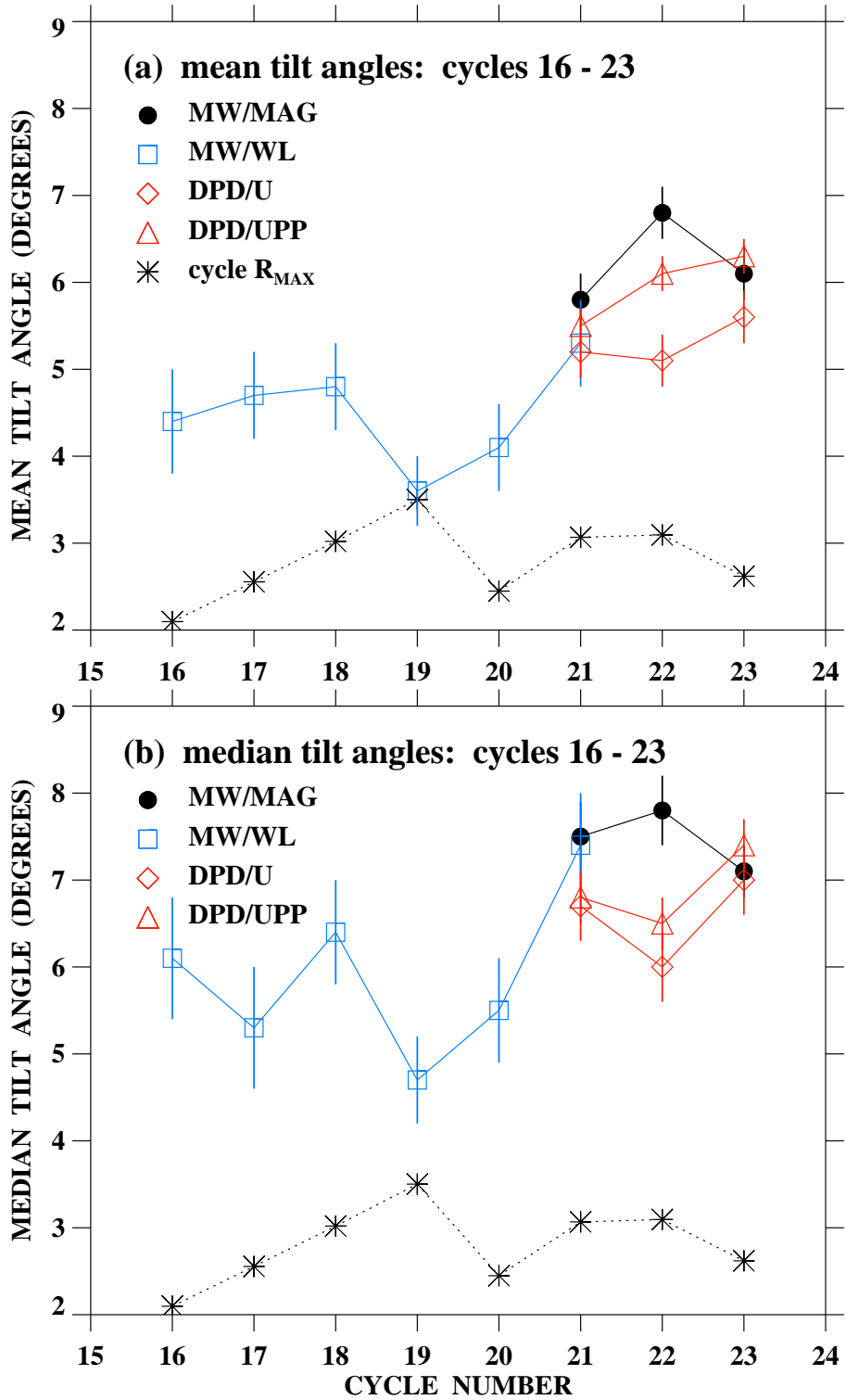}}
\vspace{-1.0cm}
\caption{Cycle-averaged tilt angles derived from MW/MAG, MW/WL, DPD/U, 
and DPD/UPP measurements.  (a) Mean values $\gamma_{\rm mean}$.  
(b) Median values $\gamma_{\rm med}$.}
\end{figure}

\newpage
\begin{figure}
\vspace{-2.0cm}
\centerline{\includegraphics[width=40pc]{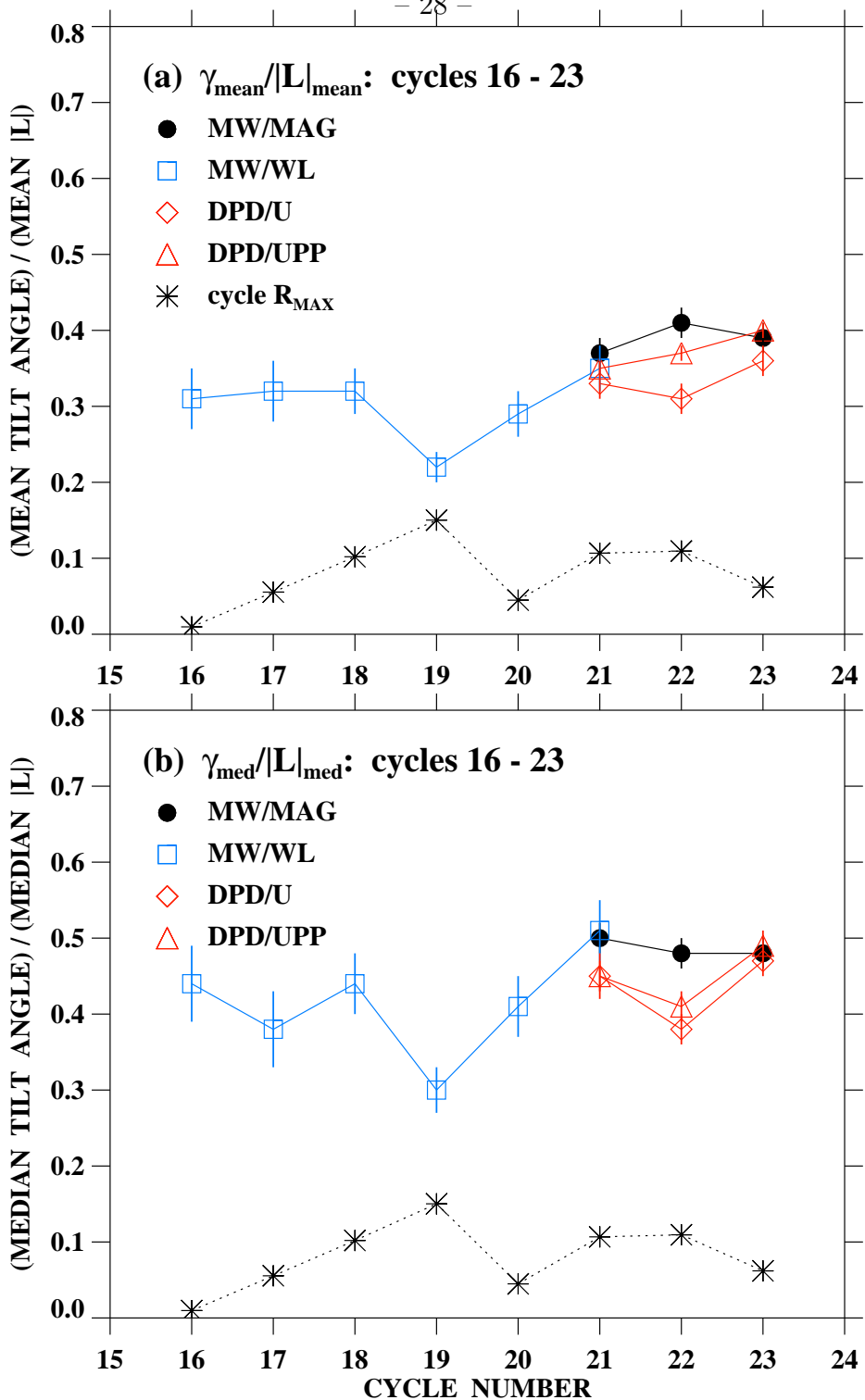}}
\vspace{-1.0cm}
\caption{Ratio of cycle-averaged tilt angle to cycle-averaged latitude, 
derived from MW/MAG, MW/WL, DPD/U, and DPD/UPP measurements.  
(a) $\gamma_{\rm mean}/\vert L\vert_{\rm mean}$.  
(b) $\gamma_{\rm med}/\vert L\vert_{\rm med}$.  Normalizing by latitude 
reduces both the cycle-to-cycle variation and the spread between 
the different data sets.}
\end{figure}

\newpage
\begin{figure}
\vspace{-5.5cm}
\centerline{\includegraphics[width=40pc]{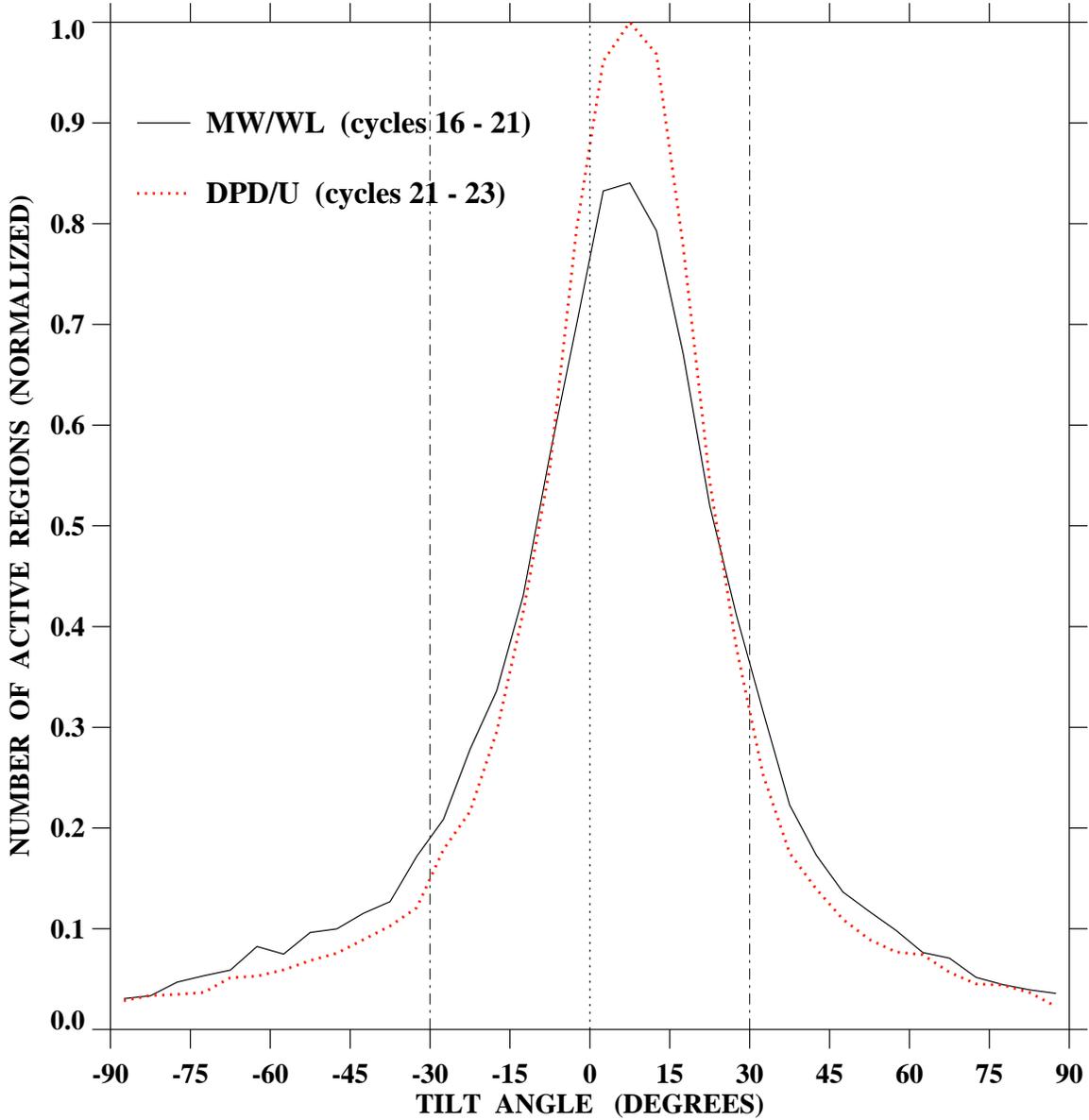}}
\vspace{-2.0cm}
\caption{Frequency distribution of MW/WL tilt angles measured for 
cycles~16--21 (black solid curve), displayed together with the 
corresponding DPD/U histogram for cycles~21--23 (red dotted curve).  
Both histograms (bin width 5$^\circ$) have been normalized by dividing by 
the total number of measurements in each data set and multiplying 
by an arbitrary constant.  The MW/WL distribution has HWHM = 17{\fdg}9, 
and shows a larger frequency of both positive and negative outliers 
than DPD/U (HWHM = 15{\fdg}4).}
\end{figure}

\newpage
\begin{figure}
\vspace{-5.0cm}
\centerline{\includegraphics[width=40pc]{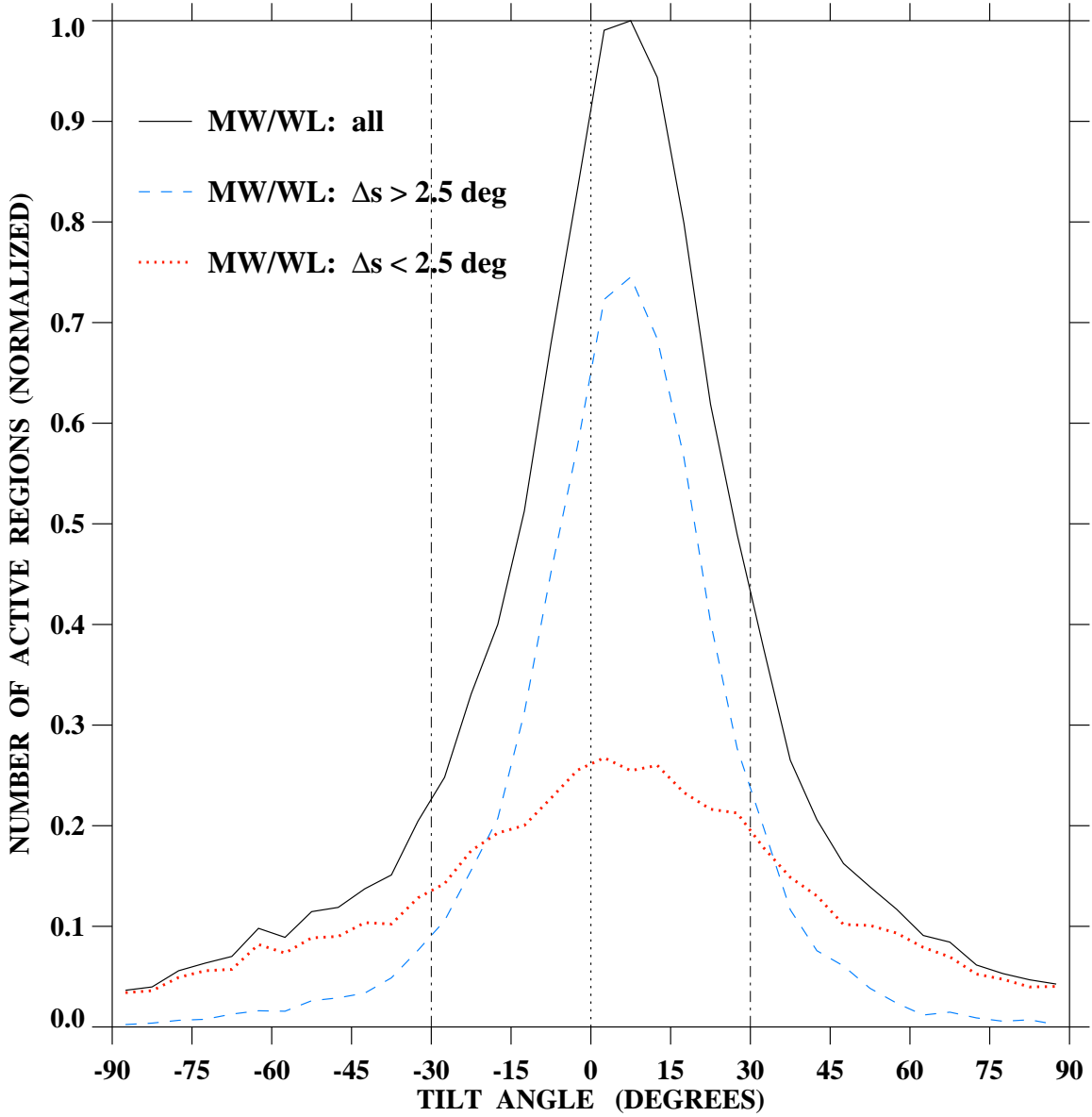}}
\vspace{-2.0cm}
\caption{The histogram of MW/WL tilt angles measured for cycles~16--21 
(black solid curve) is here separated into a component consisting of 
12,777 sunspot groups with angular separations $\Delta s > 2{\fdg}5$ 
(blue dashed curve), and a component consisting of the remaining 
9756 sunspot groups with $\Delta s < 2{\fdg}5$ (red dotted curve).  
All three histograms have bin width 5$^\circ$ and have been normalized 
by dividing by the total number of MW/WL measurements (22,533) 
and multiplying by a constant scaling factor.  The tilt angle distribution 
of sunspot groups with $\Delta s < 2{\fdg}5$, the great majority of 
which have leading and following sectors of the same polarity, 
is extremely broad (HWHM = 25{\fdg}7).  Filtering out this component 
reduces the HWHM of the original distribution from 17{\fdg}9 
to 16{\fdg}6, and removes most of the outliers in the far wings.}
\end{figure}

\newpage
\begin{figure}
\vspace{-2.5cm}
\centerline{\includegraphics[width=40pc]{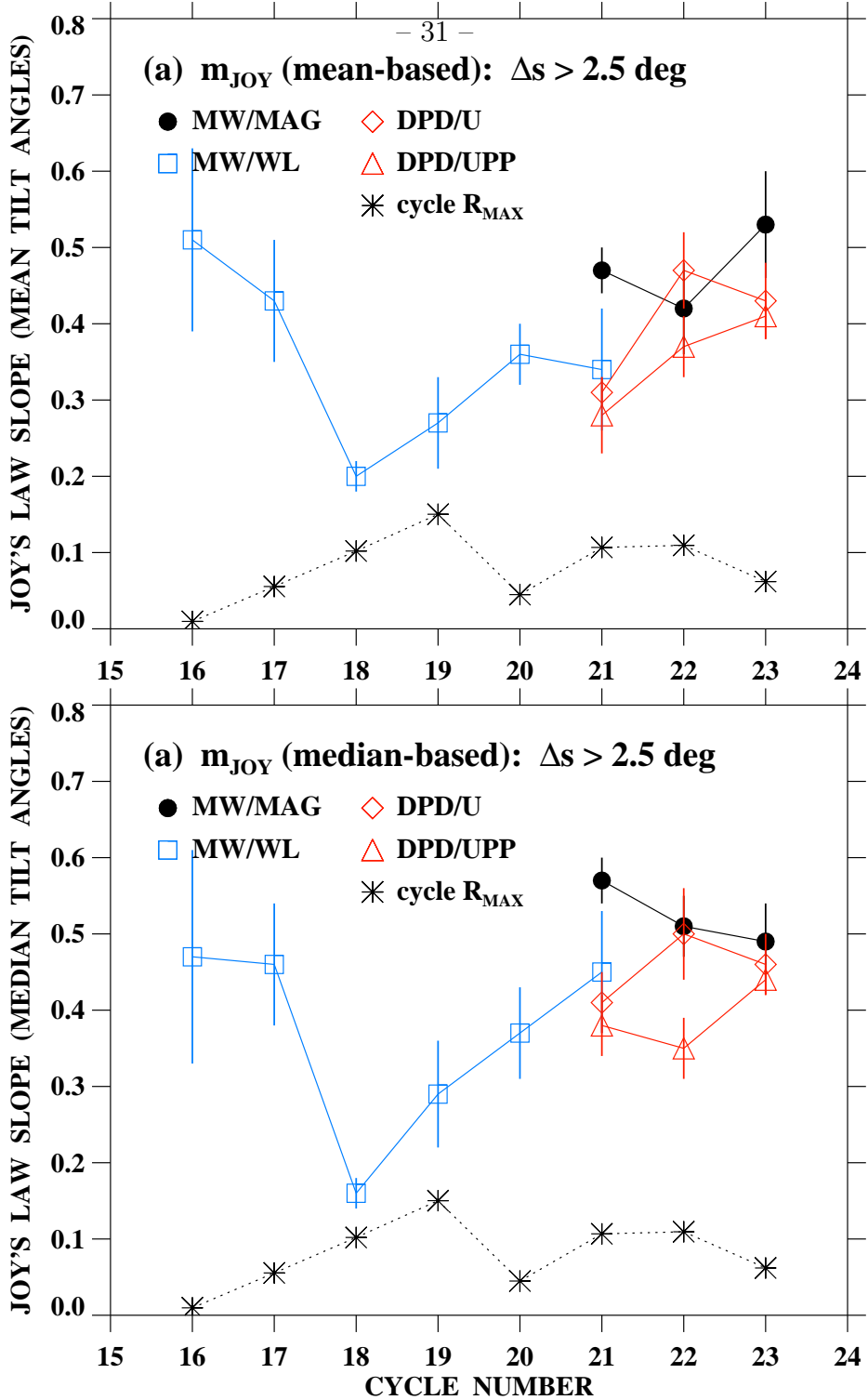}}
\vspace{-1.0cm}
\caption{Cycle-averaged Joy's law slopes derived from MW/MAG, MW/WL, 
DPD/U, and DPD/UPP measurements, after omitting active regions with 
angular separations $\Delta s < 2{\fdg}5$.  (a) Mean values.  
(b) Median values.  Compare Figure~4, where all active regions 
are included in the averages.  The filtering condition removes most 
of the ``unipolar'' sunspot groups from the MW/WL and DPD data sets.}
\end{figure}

\newpage
\begin{figure}
\vspace{-2.4cm}
\centerline{\includegraphics[width=14pc]{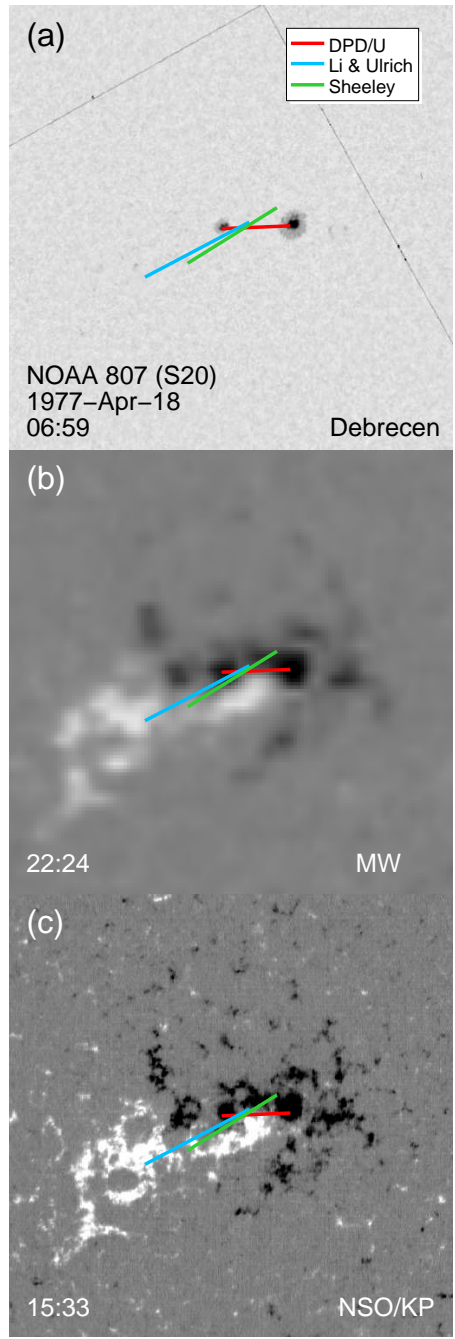}}
\vspace{0.0cm}
\caption{Comparison of white-light and magnetic tilt-angle measurements
for NOAA 807, centered at $L\simeq -20^\circ$.  (a) White-light image
taken by the Debrecen Observatory at 06:59~UT on 1977 April~18.
(b) MW magnetogram recorded at 22:24~UT.  (c) NSO/KP magnetogram
recorded at 15:33~UT.  Colored lines overplotted on each image
connect the measured centroids (corrected for differential rotation)
of the leading and following sectors.  Red: DPD/U ($\gamma = 2{\fdg}5$).
Blue: MW/MAG ($\gamma = 29{\fdg}8$; see Li \& Ulrich 2012).
Green: KP/MAG ($\gamma = 29{\fdg}7$; see Wang \& Sheeley 1989).
The white-light image is dominated by a pair of sunspots separated by 
4{\fdg}5; DPD/U assigns the easternmost spot to the following sector, 
even though the magnetograms show that it has the leading polarity.  
The resulting $\gamma$ is much smaller than that obtained from 
the magnetic measurements.  This active region also provides an example 
of longitudinal overlap between leading- and trailing-polarity sectors, 
another major source of error in white-light tilt-angle determinations 
(see also Figure~12).}
\end{figure}

\newpage
\begin{figure}
\vspace{-2.2cm}
\centerline{\includegraphics[width=15pc]{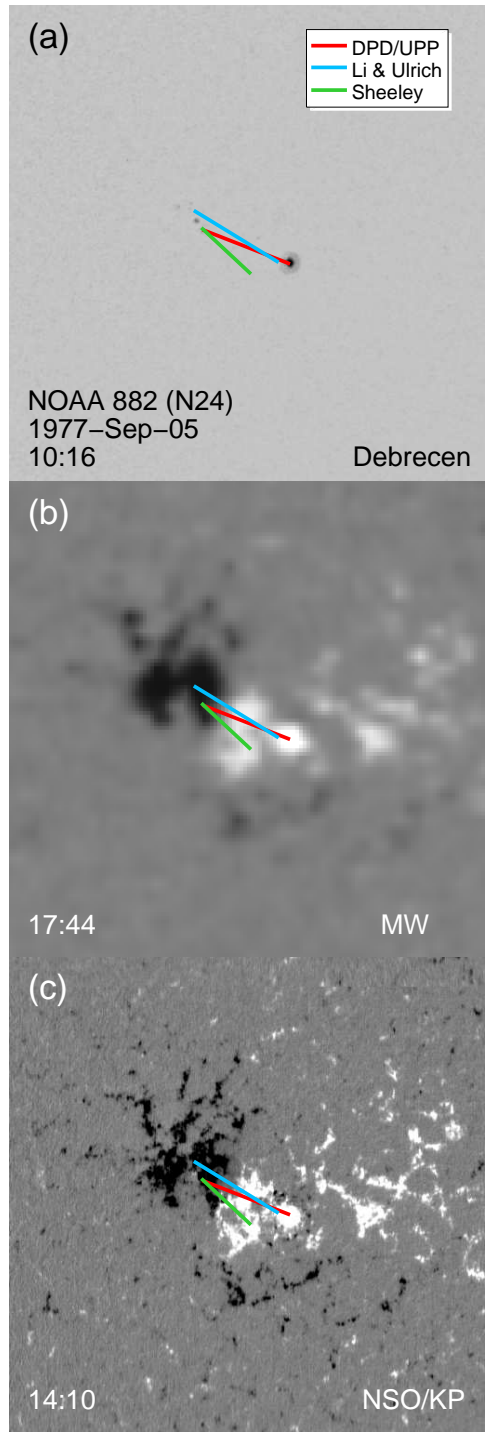}}
\vspace{0.0cm}
\caption{Comparison of white-light and magnetic tilt-angle measurements 
for NOAA 882, centered at $L\simeq +24^\circ$.  (a) White-light image 
taken by the Debrecen Observatory at 10:16~UT on 1977 September~5.  
(b) MW magnetogram recorded at 17:44~UT.  (c) NSO/KP magnetogram  
recorded at 14:10~UT.  Red line: DPD/UPP ($\gamma = 19{\fdg}0$).  
Blue line: MW/MAG ($\gamma = 29{\fdg}3$).  
Green line: KP/MAG ($\gamma = 39{\fdg}5$).  The magnetic measurements 
give larger tilt angles than DPD/UPP because the axis defined by 
the plage areas is steeper than that defined by the sunspots.  
The DPD/U measurement (not plotted) has the centroid of the 
following sector coinciding with a very small satellite umbra 
within the leading sunspot, giving a tilt angle of 12{\fdg}3.}
\end{figure}

\newpage
\begin{figure}
\vspace{-2.2cm}
\centerline{\includegraphics[width=15pc]{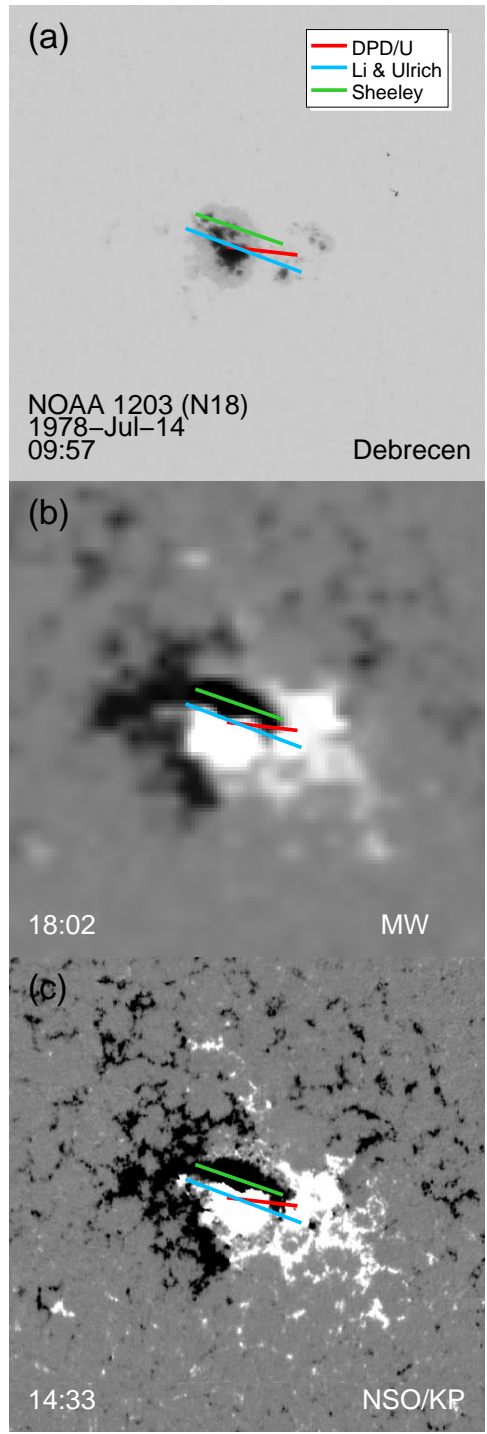}}
\vspace{0.2cm}
\caption{Comparison of white-light and magnetic tilt-angle measurements 
for NOAA 1203, centered at $L\simeq +18^\circ$.  (a) White-light image 
taken by the Debrecen Observatory at 09:57~UT on 1978 July~14.  
(b) MW magnetogram recorded at 18:02~UT.  (c) NSO/KP magnetogram  
recorded at 14:33~UT.  Red line: DPD/U ($\gamma = 6{\fdg}4$).  
Blue line: MW/MAG ($\gamma = 21{\fdg}0$).  
Green line: KP/MAG ($\gamma = 19{\fdg}4$).  In this case, 
the white-light measurements have the following sector centered on 
the large, eastern sunspot structure, whereas the magnetograms show 
that the equatorward half of this structure actually has leading polarity.  
The magnetic axis thus has a larger inclination than suggested by 
the white-light image.}
\end{figure}

\newpage
\begin{figure}
\vspace{-1.0cm}
\centerline{\includegraphics[width=18pc]{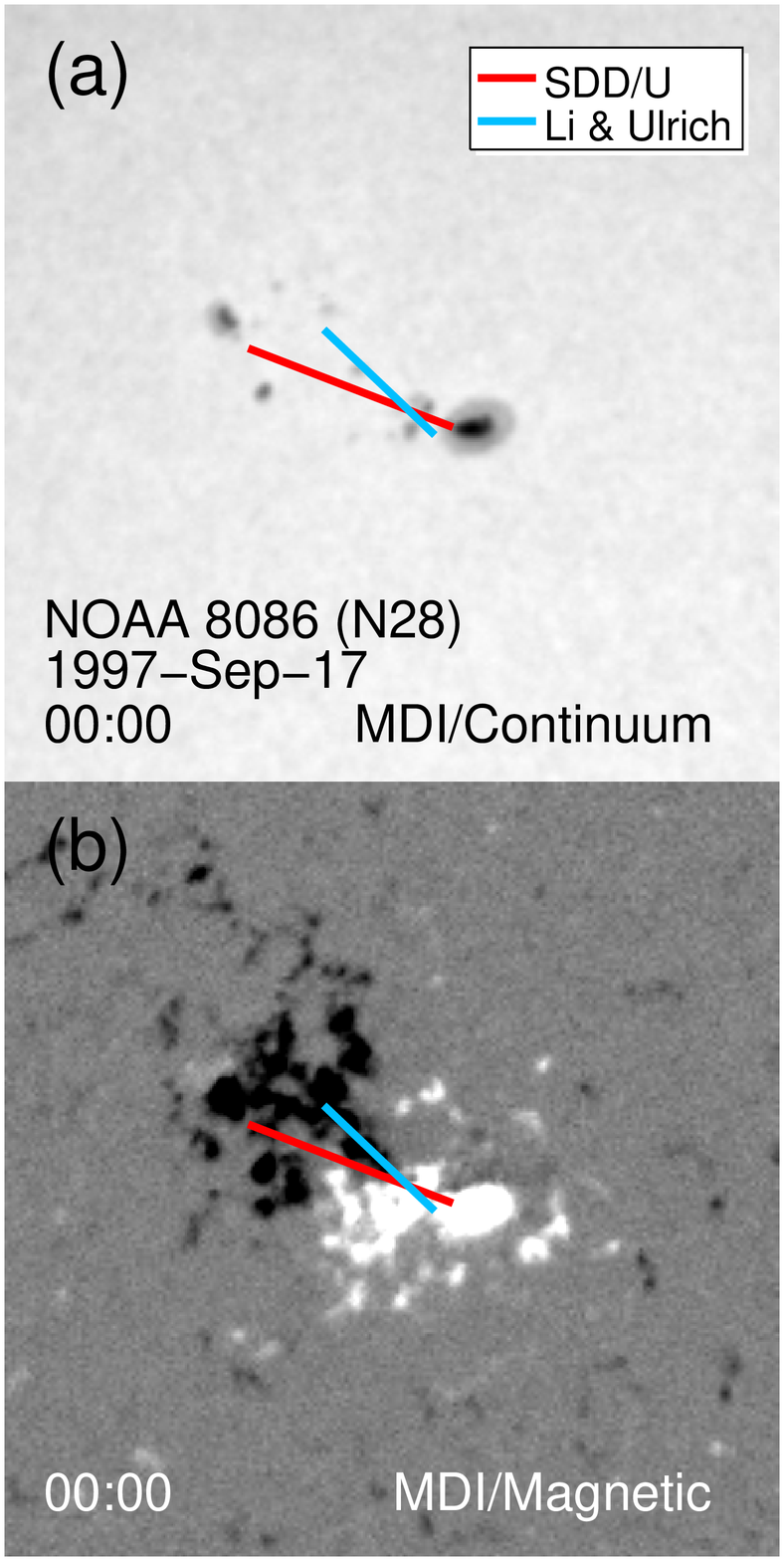}}
\vspace{0.5cm}
\caption{Comparison of white-light and magnetic tilt-angle measurements
for NOAA 8086, centered at $L\simeq +28^\circ$.  (a) MDI continuum image
recorded at 00:00~UT on 1997 September~17.  (b) Simultaneous MDI 
magnetogram.  Red line: SDD/U ($\gamma = 19{\fdg}55$).  
Blue line: MDI/MAG ($\gamma = 40{\fdg}8$; Li \& Ulrich 2012).  
Again, the magnetic tilt angle is larger because the plage areas 
visible in the magnetogram have a more steeply inclined axis 
than the sunspots seen in the continuum image.}
\end{figure}


\newpage

\begin{references}

\reference{}Babcock, H. D. 1959, ApJ, 130, 364

\reference{}Baranyi, T. 2014, MNRAS, submitted

\reference{}Baumann, I., Schmitt, D., Sch{\"u}ssler, M., \& Solanki, 
S. K. 2004, A{\&}A, 426, 1075

\reference{}Brunner, W. 1930, MiZur, 124, 67

\reference{}Caligari, P., Moreno-Insertis, F., \& Sch{\"u}ssler, M. 1995, 
ApJ, 441, 886

\reference{}Cameron, R. H., Jiang, J., Schmitt, D., \& Sch{\"u}ssler, M. 2010,
ApJ, 719, 264

\reference{}Cameron, R. H., \& Sch{\"u}ssler, M. 2012, A{\&}A, 548, A57

\reference{}Chapman, G. A., Dobias, J. J., \& Arias, T. 2011, ApJ, 728, 150

\reference{}Dasi-Espuig, M., Solanki, S. K., Krivova, N. A., Cameron, R., \& 
Pe{\~n}uela, T. 2010, A{\&}A, 518, A7

\reference{}Dasi-Espuig, M., Solanki, S. K., Krivova, N. A., Cameron, R., \& 
Pe{\~n}uela, T. 2013, A{\&}A, 556, C3

\reference{}Dikpati, M., \& Charbonneau, P. 1999, ApJ, 518, 508

\reference{}D'Silva, S., \& Choudhuri, A. R. 1993, A{\&}A, 272, 621

\reference{}Fan, Y., Fisher, G. H., \& DeLuca, E. E. 1993, ApJ, 405, 390

\reference{}Fan, Y., Fisher, G. H., \& McClymont, A. N. 1994, ApJ, 436, 907

\reference{}Fisher, G. H., Fan, Y., \& Howard, R. F. 1995, ApJ, 438, 463

\reference{}Foukal, P. 1993, SoPh, 148, 219

\reference{}Gy\H{o}ri, L., Baranyi, T., \& Ludm{\'a}ny, A. 2011, 
The Physics of Sun and Star Spots (IAU Symp. 273), ed. D. P. Choudhary \& 
K. G. Strassmeier (Cambridge: Cambridge Univ. Press), 403

\reference{}Hale, G. E., Ellerman, F., Nicholson, S. B., \& Joy, A. H. 1919, 
ApJ, 49, 153

\reference{}Hathaway, D. H., \& Upton, L. 2014, JGR, 119, 3316

\reference{}Howard, R. F. 1989, SoPh, 123, 271

\reference{}Howard, R. F. 1991a, SoPh, 132, 49

\reference{}Howard, R. F. 1991b, SoPh, 136, 251

\reference{}Howard, R. F. 1996, SoPh, 169, 293

\reference{}Howard, R., Gilman, P. A., \& Gilman, P. I. 1984, ApJ, 283, 373

\reference{}Ivanov, V. G. 2012, Ge{\&}Ae, 52, 999

\reference{}Jiang, J., Cameron, R. H., Schmitt, D., \& I{\c s}ik, E. 2013, 
A{\&}A, 553, A128

\reference{}Jiang, J., Cameron, R. H., Schmitt, D., \& 
Sch{\"u}ssler, M. 2011a, A{\&}A, 528, A82

\reference{}Jiang, J., Cameron, R. H., Schmitt, D., \& 
Sch{\"u}ssler, M. 2011b, A{\&}A, 528, A83

\reference{}Jiang, J., Cameron, R. H., \& Sch{\"u}ssler, M. 2014, 
ApJ, 791, 5

\reference{}Kitchatinov, L. L., \& Olemskoy, S. V. 2011, AstL, 37, 656

\reference{}Kosovichev, A. G., \& Stenflo, J. O. 2008, ApJ, 688, L115

\reference{}Li, J., \& Ulrich, R. K. 2012, ApJ, 758, 115

\reference{}Mackay, D. H., Priest, E. R., \& Lockwood, M. 2002, SoPh, 207, 291

\reference{}McClintock, B. H., \& Norton, A. A. 2013, SoPh, 287, 215

\reference{}Mu{\~n}oz-Jaramillo, A., Dasi-Espuig, M., Balmaceda, L. A., \& 
DeLuca, E. E. 2013, ApJ, 767, L25

\reference{}Mu{\~n}oz-Jaramillo, A., Nandy, D., Martens, P. C. H., \& 
Yeates, A. R. 2010, ApJ, 720, L20

\reference{}Murak{\"o}zy, J., Baranyi, T., \& Ludm{\'a}ny, A. 2014, SoPh, 
289, 563

\reference{}Schrijver, C. J. 1987, A{\&}A, 180, 241

\reference{}Sch{\"u}ssler, M., \& Baumann, I. 2006, A{\&}A, 459, 945

\reference{}Sheeley, N. R., Jr. 1966, ApJ, 144, 723

\reference{}Sheeley, N. R., Jr. 2008, ApJ, 680, 1553

\reference{}Sheeley, N. R., Jr., Cooper, T. J., \& Anderson, J. R. L. 2011, 
ApJ, 730, 51

\reference{}Stenflo, J. O., \& Kosovichev, A. G. 2012, ApJ, 745, 129

\reference{}Tian, L., Liu, Y., \& Wang, H. 2003, SoPh, 215, 281

\reference{}Tlatov, A. G., Tlatova, K. A., Vasil'eva, V. V., Pevtsov, A. A., 
\& Mursula, K. 2014, AdSpR, in press

\reference{}Upton, L., \& Hathaway, D. H. 2014a, ApJ, 780, 5

\reference{}Upton, L., \& Hathaway, D. H. 2014b, ApJ, 792, 142

\reference{}Wang, Y.-M., \& Sheeley, N. R., Jr. 1989, SoPh, 124, 81

\reference{}Wang, Y.-M., \& Sheeley, N. R., Jr. 1991, ApJ, 375, 761

\end{references}
\end{document}